\documentclass{article}
\usepackage{authblk}
\usepackage{geometry}
\usepackage{graphicx}
\usepackage[colorlinks=true]{hyperref}
\usepackage{xcolor}
\usepackage{amsfonts}
\usepackage{amsmath}
\usepackage{amssymb}
\usepackage[most]{tcolorbox}
\usepackage{booktabs}
\usepackage{array}
\usepackage{tikz}
\usetikzlibrary{positioning,decorations.pathreplacing,arrows,arrows.meta,fit,shapes,shapes,arrows,shapes.multipart,calc,backgrounds, decorations.markings}
\usepackage{pgfplots}

\DeclareMathOperator*{\argmin}{arg\,min}
\newcommand{\bs}{\boldsymbol}

\newtcolorbox[auto counter]{mybox}[2][]{
    float, enhanced, 
    colback=blue!5, colframe=red!60, boxrule=0.6pt, arc=1.5mm, 
    left=1mm, right=1mm, top=0.5mm, bottom=0.5mm, 
    title=\textbf{Box~\thetcbcounter\ |~#2},
    fonttitle=\bfseries, coltitle=black,
    label=#1
}

\title{Enabling risk and reliability analysis in next-generation computational tools via multifidelity methods}
\author{Anirban Chaudhuri}
\author{Karen Willcox}
\affil{University of Texas at Austin, TX 78712, USA}
\date{}

\begin{document}
\pgfmathdeclarefunction{frechet}{3}{%
	\pgfmathparse{(#1/#2)*((x-#3)/#2)^(-1-#1)*exp(-((x-#3)/#2)^(-#1))}%
}

\maketitle
\begin{abstract}
Risk and reliability analysis are fundamental to the design, assessment, and operation of safety-critical complex systems; however, their integration into next-generation computational tools is constrained by the cost of uncertainty quantification with expensive computational models. The computational challenge is particularly prohibitive for tail-sensitive quantities, such as probabilities of failure, conditional values-at-risk, and buffered probabilities of failure, where accurate estimation can require a large number of model evaluations. Multifidelity methods provide a framework for addressing this computational bottleneck by combining information from models of varying accuracy and cost while retaining the predictive accuracy of the high-fidelity model. This review summarizes the state of the art in methods that enable scalable risk and reliability analysis using multifidelity control variates, multifidelity importance sampling, and multifidelity active learning.
\end{abstract}

\section{Introduction}

Risk assessment and reliability analysis are critical aspects of engineering and scientific workflows, playing particularly important roles in system analysis, assessment, certification, system design, experimental design, and decision under uncertainty. The emergence of powerful next-generation computational tools such as digital twins~\cite{tuegel2011digitaltwin,glaessgen2012digital,thelen2022review} makes these workflows increasingly computational, yet there is a gap in capabilities to achieve quantitative risk and reliability assessments at scale~\cite{nasem2023digitaltwins}. For example, AI-enabled and simulation-based design of aerospace systems and energy infrastructure requires statistical analysis to estimate tail risks and rare-event probabilities using computational models that are increasingly expensive to evaluate~\cite{karve2020digital,kapteyn2021pgm,ferrari2024mechanicalaerospace}. Similarly, AI-enabled digital twins that support medical decision-making must go beyond point estimates; they must provide the decision-maker with characterization and quantification of uncertainties~\cite{sel2025survey,laubenbacher2024digital}. Over the past two decades a rich set of uncertainty quantification methods have emerged to address these needs (see e.g.,~\cite{smith2024}), but it remains a challenge to achieve computationally tractable uncertainty quantification, especially when the digital twin embeds expensive high-fidelity simulations that are needed to characterize low-probability, high-consequence events. This computational challenge is being addressed by the recent advancements in multifidelity uncertainty quantification methods.

Multifidelity uncertainty quantification methods provide a structured way to assess risk and reliability at acceptable computational cost without compromising accuracy. Multifidelity methods leverage multiple computational models of varying degrees of fidelity and computational cost, including detailed high-fidelity models, reduced-order and surrogate models, reduced-resolution models, data-driven models, and simplified-physics models. This review examines multifidelity methods that accelerate risk and reliability analysis while maintaining rigorous error control, and discusses their integration into next-generation computational tools such as digital twins.

\begin{figure}[!hbt]
    \centering
    \includegraphics[width=\textwidth]{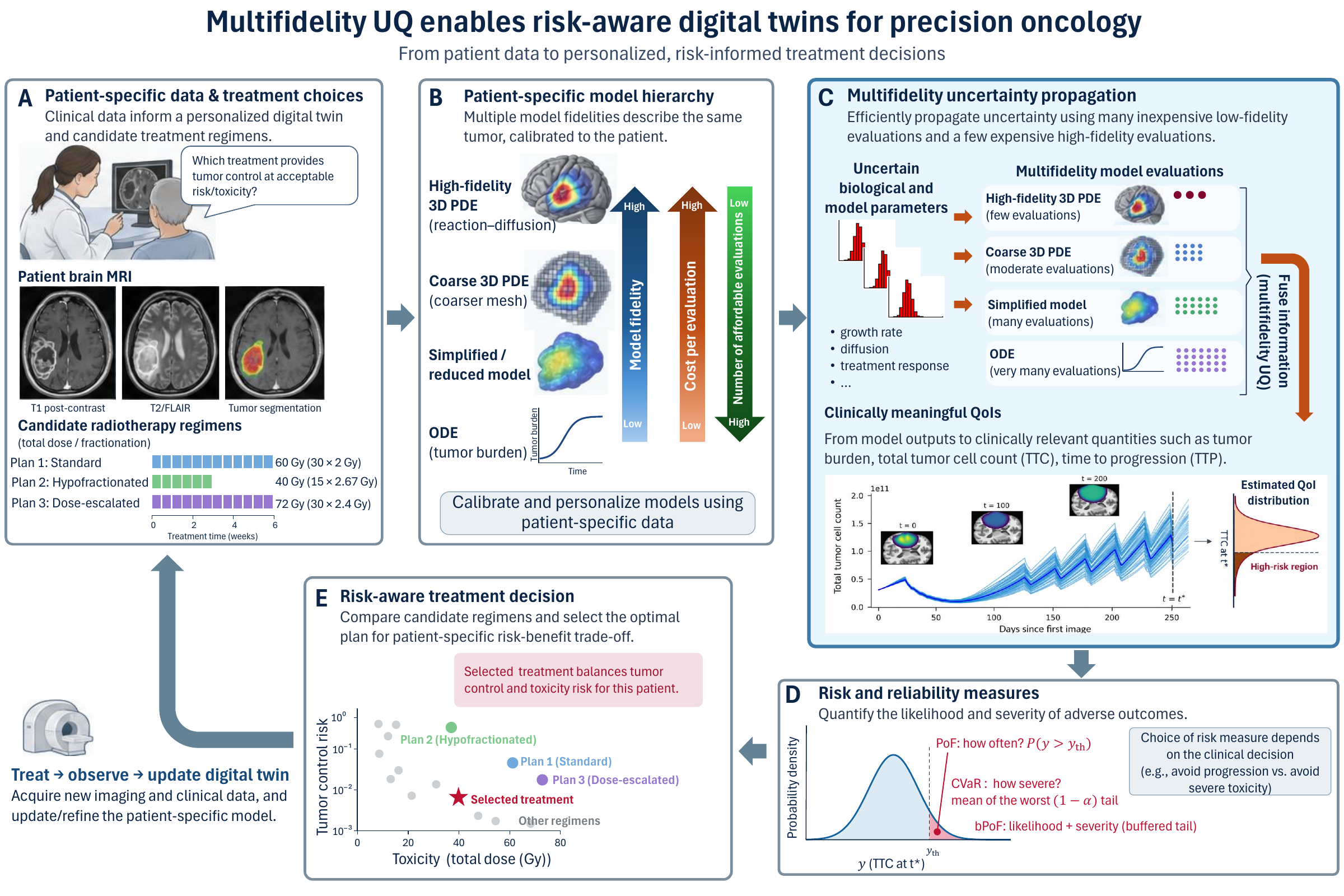}
    \caption{Risk-aware digital twin framework for precision oncology using multifidelity methods. (A) Patient-specific data and candidate treatment regimens define the clinical decision problem. (B) These data are used to personalize a hierarchy of tumor growth models with varying fidelities and computational costs. (C) Multifidelity methods efficiently propagate uncertainty across these models to estimate predictive distributions of clinically meaningful QoIs, (D) These distributions are summarized using risk and reliability measures. (E) Candidate treatments are compared on the basis of tumor control risk and toxicity trade-offs. The digital twin is iteratively updated as new patient data become available.}
    \label{fig:oncorisk}
\end{figure}

As one concrete example, Figure~\ref{fig:oncorisk} illustrates the risk-aware digital twin paradigm for precision oncology, in which patient-specific data calibrate a hierarchy of tumor growth models, multifidelity uncertainty quantification is used to estimate distributions of clinically meaningful quantities of interest (QoIs), and risk or reliability measures inform treatment selection. 
As described in~\cite{chaudhuripredictive2023}, incorporating uncertainty quantification into medical digital twins is essential to account for uncertainty in the digital twin models by continuous integration of incoming patient data, to forecast the parameter uncertainty to estimate risk associated with the therapeutic outcomes, and to support risk-aware clinical decision-making under uncertainty. 
Multifidelity methods play a key role in making uncertainty-aware, patient-specific decision support computationally tractable. Digital twin frameworks across medicine and engineering share a common need to integrate models and data under uncertainty, making multifidelity methods especially attractive for efficient prediction, risk assessment, and operational decision-making.

Section~\ref{s:risk} introduces common risk and reliability measures---probability of failure, conditional value-at-risk (CVaR), and buffered probability of failure---and discusses their important role in engineering and scientific workflows. Section~\ref{s:mfrisk} then presents the multifidelity setting and discusses three specific classes of multifidelity methods for improving estimator efficiency for these risk and reliability measures: multifidelity control variates, multifidelity importance sampling, and multifidelity active learning with probabilistic surrogates. Finally, Section~\ref{s:outlook} concludes with an outlook and discussion of how these methods can enable next-generation computational tools with scalable reliability and risk assessment.

\section{Risk and reliability measures} \label{s:risk}

Risk and reliability measures quantify how a system responds to uncertain inputs. Let $\bs{\xi}\in\Xi$ denote a vector of uncertain inputs with probability density $\pi(\bs{\xi})$.
A computational model maps these uncertain inputs to an output QoI as
\begin{equation}
y = f(\bs{\xi}),
\label{eq:qoi}
\end{equation}
where $f:\Xi\rightarrow\mathbb{R}$ represents the computational simulation mapping from input space $\Xi$ to a scalar QoI $y \in \mathbb{R}$, such as a structural response, temperature, stress, energy output, or treatment response. 
The uncertainty in $\bs{\xi}$ induces uncertainty in the QoI, and so in (\ref{eq:qoi}) $y$ is a random variable.

The mean QoI response is given by 
$\mu = \mathbb{E}_{\pi}[y] = \int_{\Xi} y \pi(\bs{\xi}) \, \mathrm{d}\bs{\xi}$ 
and the variance by
$\sigma^2 = \mathbb{V}_{\pi}[y] = \int_{\Xi} \left( y - \mu \right)^2 \pi(\bs{\xi}) \, \mathrm{d}\bs{\xi}$. 
Box~\ref{box:risk} shows four commonly used risk and reliability measures that are computed by estimating different statistics of the random variable $y$. 
Risk and reliability analysis focus on the tails of the distribution, and are typically used when the tails correspond to undesirable or unsafe system behavior. Let $y_{\mathrm{th}}$ denote a prescribed failure threshold, with failure defined by $y > y_{\mathrm{th}}$. Then the probability of failure (PoF) is 
\begin{equation}
\label{eq:pof}
p_f = \mathbb{P}_{\pi} \left(y>y_{\mathrm{th}} \right)
= \int_{\Xi} \mathbb{I} \left[ y>y_{\mathrm{th}} \right] \pi(\bs{\xi}) \, \mathrm{d}\bs{\xi} =\mathbb{E}_{\pi}\left[\mathbb{I}\left[y>y_{\mathrm{th}}\right]\right],
\end{equation}
where $\mathbb{I}[\cdot]$ is the indicator function. Equivalently, if $F_y$ denotes the cumulative distribution function of $y$, then $p_f = 1-F_y(y_{\mathrm{th}})$. The PoF is a popular measure of reliability because it directly quantifies the likelihood of violating a prescribed performance threshold. However, the indicator function in Eq.~\eqref{eq:pof} does not distinguish between outcomes that barely violate the threshold and outcomes that substantially exceed it. This distinction between the likelihood and severity of exceedances motivates tail-sensitive risk measures.
\begin{mybox}[box:risk]{Key concepts in risk and reliability analysis}

\renewcommand{\arraystretch}{1.15}
\begin{tabular}{>{\bfseries}p{0.17\linewidth} p{0.77\linewidth}}
\toprule
\textbf{Term} & \textbf{Meaning} \\
\midrule
Quantity of interest (QoI) &
A model output used to characterize system behaviour. \\

Probability of failure (PoF) &
The probability that the QoI exceeds a prescribed failure threshold, estimating \textit{how often the threshold is exceeded}. \\ 

Value-at-risk (VaR) &
The $\alpha$-quantile of a QoI distribution. It identifies a value below which a fraction $\alpha$ of outcomes lie, indicating \textit{where the tail begins}. \\

Conditional value-at-risk (CVaR) &
A tail-risk measure that quantifies the average severity of outcomes in the upper tail beyond a specified quantile. It estimates \textit{how severe the outcomes are, on average, in the tail}. It is also known as the superquantile or expected shortfall. \\

Buffered probability of failure (bPoF) &
A reliability measure based on the superquantile that accounts for both the likelihood and severity of threshold exceedances. It provides a conservative alternative to the conventional probability of failure and estimates \textit{what probability level corresponds to a tail average reaching the failure threshold}. \\
\bottomrule
\end{tabular}

\vspace{-1mm}
\begin{center}
\begin{tikzpicture}[scale=1,auto]

\begin{axis}[
    xshift=-4cm,
    xmin=0.3, xmax=3.5, ymin=0, ymax=1.2,
    height=3.5cm, width=5.0cm,
    black!60, very thick,
    domain=0.3:3, samples=100,
    axis lines=left,
    every axis y label/.style={anchor=south west, rotate=90, align=flush center, inner sep=0.3ex, font=\small},
    ylabel={Probability\\[-0.1cm] density},
    every axis x label/.style={at=(current axis.right of origin), anchor=north, font=\small},
    xlabel={$y$}, ytick=\empty, xtick={2.1}, xticklabels={$y_{\mathrm{th}}$},
    every tick/.style={black,very thick}, tick label style={font=\small,black},
    enlargelimits=false, clip=true, axis on top, smooth
]

\addplot[
    fill=red!60,
    draw=none,
    domain=2.1:3
] {frechet(3,1.5,0)} \closedcycle;

\addplot[very thick,black!80, domain=0.3:3] {frechet(3,1.5,0)};

\draw[decoration={markings, mark=at position 1 with {\arrow[scale=0.5]{*}}},
    postaction={decorate},
    black!70
]
(axis cs:2.65,0.78)
node [black,above,xshift=-0.45cm]
{$p_f=\mathbb{P}_{\pi}\left(y>y_{\mathrm{th}}\right)$}
-- (axis cs:2.5,0.07);

\end{axis}

\begin{axis}[
    xshift=1.25cm,
    xmin=0.3, xmax=3.5, ymin=0, ymax=1.2,
    height=3.5cm, width=5.0cm,
    black!60, very thick,
    domain=0.3:3, samples=100,
    axis lines=left,
    every axis y label/.style={anchor=south west,rotate=90,
        align=flush center,inner sep=0.3ex,font=\small},
    ylabel={Probability\\[-0.1cm] density},
    every axis x label/.style={at=(current axis.right of origin),
        anchor=north, font=\small},
    xlabel={$y$}, ytick=\empty, xtick={1.35,2.1}, 
    xticklabels={
        {\hspace{-1mm}$\operatorname{VaR}_{\alpha}$},
        {$\quad \operatorname{CVaR}_{\alpha}$}
    },
    every tick/.style={black,very thick}, tick label style={font=\small,black},
    enlargelimits=false, clip=true, axis on top, smooth
]

\addplot[fill=red!30, draw=none, domain=0.3:1.35] {frechet(3,1,0)} \closedcycle;
\addplot[fill=blue!65, draw=none, domain=1.35:3] {frechet(3,1,0)} \closedcycle;
\addplot[very thick, black!80, domain=0.3:3] {frechet(3,1,0)};

\node at (axis cs:1.0,0.30)
    [fill=white,font=\footnotesize,inner sep=0.3ex]
    {$\alpha$};

\node at (axis cs:1.8,0.14)
    [fill=white,font=\footnotesize,inner sep=0.1ex]
    {$1-\alpha$};

\end{axis}

\begin{axis}[
    xshift=6.3cm, xmin=0.3, xmax=3.5, ymin=0, ymax=1.2,
    height=3.5cm, width=5.0cm,
    black!60, very thick,
    domain=0.3:3, samples=100,
    axis lines=left,
    every axis y label/.style={anchor=south west,rotate=90,
        align=flush center,inner sep=0.3ex,font=\small},
    ylabel={Probability\\[-0.1cm] density},
    every axis x label/.style={at=(current axis.right of origin),
        anchor=north, font=\small},
    xlabel={$y$}, ytick=\empty, xtick={1.45,2.2},
    xticklabels={
        {\hspace{-1mm}$\operatorname{VaR}_{\alpha^\star}$},
        {\shortstack{$\qquad \operatorname{CVaR}_{\alpha^\star}$\\[-2pt]$\;=y_{\mathrm{th}}$}}
    },
    every tick/.style={black,very thick},
    tick label style={font=\small, black, align=center},
    enlargelimits=false, clip=true, axis on top, smooth
]

\addplot[fill=blue!55, draw=none, domain=1.45:3
] {frechet(3,1.25,0)} \closedcycle;

\addplot[fill=red!60, draw=none, domain=2.2:3
] {frechet(3,1.25,0)} \closedcycle;

\addplot[very thick, black!80, domain=0.3:3
] {frechet(3,1.25,0)};

\node at (axis cs:1.78,0.13)
    [fill=white,font=\footnotesize,inner sep=0.1ex]
    {buffer};

\draw[decoration={markings, mark=at position 1 with {\arrow[scale=0.5]{*}}},
    postaction={decorate},
    black!70
]
(axis cs:2.65,0.45)
node [black,above,yshift=-0.1cm]
{$p_f$}
-- (axis cs:2.4,0.03);

\draw[decoration={markings, mark=at position 1 with {\arrow[scale=0.5]{*}}}, postaction={decorate}, black!70]
(axis cs:2.05,0.78)
node [black,above,xshift=0.2cm,yshift=-0.1cm]
{$\bar p_f=1-\alpha^\star$}
-- (axis cs:1.65,0.18);

\end{axis}

\node[font=\small\bfseries] at (-2,-0.95) {(a) PoF};
\node[font=\small\bfseries] at (3,-0.95) {(b) CVaR};
\node[font=\small\bfseries] at (8,-0.95) {(c) buffered PoF};

\end{tikzpicture}
\end{center}
\end{mybox}

Tail-sensitive risk measures account for the magnitude of unsafe outcomes. For a given distribution of $y$, the value-at-risk (VaR) is defined as the $\alpha$-quantile, 
\begin{equation}
\label{eq:var}
\operatorname{VaR}_{\alpha}(y) = \inf \left\{y\in\mathbb{R}:F_y(y)\geq\alpha \right\}=F_y^{-1}(\alpha), \qquad 0<\alpha<1.
\end{equation}
The $\alpha$-quantile identifies the threshold below which a fraction $\alpha$ of the outcomes lie, with the remaining fraction $1-\alpha$ constituting the upper tail of the distribution, as illustrated in Box~\ref{box:risk}. The conditional value-at-risk (CVaR) characterizes the average severity of outcomes in the upper $(1-\alpha)$ tail~\cite{rockafellar2000optimization,RTRockafellar_SUryasev_2002a,rockafellar2013superquantiles,RTRockafellar_JORoyset_2015a}. CVaR can be expressed in terms of VaR as
\begin{equation}
\label{eq:cvar_quantile}
\operatorname{CVaR}_{\alpha}(y) = \frac{1}{1-\alpha} \int_{\alpha}^{1} \operatorname{VaR}_{\gamma}(y) \, \mathrm{d}\gamma.
\end{equation}
For a continuous distribution of $y$, this corresponds to the expected value of $y$ in the upper $(1-\alpha)$ tail, given by $\mathbb{E}_\pi [y \mid y \ge \operatorname{VaR}_{\alpha}(y)]$, and thus estimates the average severity of outcomes in this tail, as illustrated in Box~\ref{box:risk}. CVaR can be equivalently written as
\begin{equation}
\label{eq:cvar}
\operatorname{CVaR}_{\alpha}(y) 
= \operatorname{VaR}_{\alpha}(y) + \frac{1}{1-\alpha}\mathbb{E}_\pi  \left[\left( y-\operatorname{VaR}_{\alpha}(y) \right)_+ \right]
= \inf_{\gamma\in\mathbb{R}} \left\{ \gamma + \frac{1}{1-\alpha} \mathbb{E}_{\pi} \left[(y-\gamma)_+\right]\right\},
\end{equation}
where $(\cdot)_+\equiv\max(\cdot,0)$. The latter optimization-based representation in (\ref{eq:cvar}) is particularly useful computationally because it expresses CVaR in terms of an expectation and a scalar optimization problem.

A related tail-sensitive measure for reliability analysis is the buffered probability of failure (bPoF) \cite{rockafellar2010buffered} given by 
\begin{equation}
\bar{p}_f = 1-\alpha^\star, \qquad \text{where } \operatorname{CVaR}_{\alpha^\star}(y) = y_{\mathrm{th}}
\label{eq:bpof}
\end{equation}
when such an $\alpha^\star$ exists. The value $\alpha^\star$ is the quantile level at which the average response in the upper $1-\alpha^\star$ tail, measured by $\operatorname{CVaR}_{\alpha^\star}(y)$, reaches the prescribed failure threshold $y_{\mathrm{th}}$, as illustrated in Box~\ref{box:risk}. Thus, bPoF is the tail probability corresponding to the CVaR level that reaches the failure threshold. Unlike PoF, bPoF therefore accounts for both the likelihood and severity of threshold exceedances. Under typical conditions, the bPoF is at least as large as the conventional PoF, providing a more conservative characterization of reliability~\cite{rockafellar2010buffered,mafusalov2018buffered,norton2019calculating,basova2011computational,chaudhuri2022certifiable}.

For engineering and scientific problems where the input-to-QoI map takes a general nonlinear form (\ref{eq:qoi}), PoF, bPoF and CVaR are estimated using Monte Carlo simulation, which generates many realizations of the uncertain inputs, evaluates the computational model for each realization to obtain samples of the QoI using \eqref{eq:qoi}, and then estimates from the QoI samples the desired statistics of interest, such as the risk and reliability measures. For example, given $N$ independent QoI samples $\{y_i\}_{i=1}^N$, the Monte Carlo estimator of the mean is $\mathbb{E}_{\pi}[y]\approx \frac{1}{N} \sum_{i=1}^{N}y_i$. The cost of estimating these risk and reliability measures thus depends on the cost of a single model evaluation (one evaluation per Monte Carlo sample) multiplied by the number of evaluations required. This computational burden is typically severe for tail quantities and small probabilities, since Monte Carlo methods require sample sizes of thousands or more in order to accurately characterize the tails of the distribution. 
Multifidelity uncertainty quantification methods described next address this computational challenge by using cheaper models to reduce the number of expensive high-fidelity evaluations required.

\section{Multifidelity risk and reliability analysis} \label{s:mfrisk}

\subsection{Introducing the multifidelity setting} \label{s:prob}
The QoI in Eq.~\eqref{eq:qoi} is generally obtained from a high-fidelity computational model whose results are trusted by the engineer or scientist but whose evaluation can require substantial computational resources. 
Estimation of risk and reliability measures via Monte Carlo simulation requires that this high-fidelity model be evaluated thousands of times, and this typically goes well beyond the available budget. A multifidelity method seeks to use the available budget optimally by leveraging cheaper, approximate models. Different multifidelity uncertainty quantification methods use different theoretical bases for optimally distributing evaluations across multiple models and for ensuring that the resulting estimated risk and reliability measures remain accurate.

We denote the highest fidelity model by $f^{(0)}(\boldsymbol{\xi})$, where the superscript $(0)$ denotes the high-fidelity model. We consider a collection of $L$ lower-fidelity models, $\left\{f^{(1)},\ldots,f^{(L)}\right\}$. The low-fidelity models provide cheaper, but generally less accurate, approximations of the same QoI. For a model of fidelity level $\ell$, we have $y^{(\ell)} = f^{(\ell)}(\boldsymbol{\xi}), \ell=0,\ldots,L$, with corresponding computational cost $c_\ell$. We order the models so that $c_0 > c_1 \geq \cdots \geq c_L$. The models do not need to form a nested sequence or possess a strict hierarchy in accuracy except knowing the highest-fidelity model. In the context of the oncology digital twin illustrated in Figure~\ref{fig:oncorisk}, the high-fidelity model can be a fully-resolved three-dimensional parametric partial differential equation model of tumor growth. Lower-fidelity alternatives can include models with coarser spatial discretizations, lumped-parameter ordinary differential equation models, or projection-based reduced-order models, which provide varying trade-offs between computational cost and predictive accuracy while describing the same underlying tumor growth process.

The central objective of multifidelity risk and reliability analysis is to reduce the number of expensive, high-fidelity evaluations required for risk and reliability estimation by supplementing with cheaper, lower-fidelity model solves. Multifidelity methods have been shown to provide orders of magnitude reductions in overall computational cost without compromising accuracy.
An important consideration in multifidelity methods is the statistical relationship between the outputs of models at different fidelity levels. In particular, lower-fidelity models are useful when their outputs provide information about the high-fidelity response. This relationship can be quantified through correlation, which measures how strongly the outputs of two models vary together as the uncertain inputs change. For example, if a lower-fidelity model consistently predicts larger responses for the same uncertain input realizations that produce larger responses in the high-fidelity model, the two models exhibit positive correlation. Note that strong correlation does not require the models to produce identical predictions or even have small prediction errors; a lower-fidelity model can have systematic bias while still capturing the correct trends of the high-fidelity response. The multifidelity strategies differ in how lower-fidelity information is incorporated. Information from lower-fidelity models can: (i)~be used as correlated control variates (multifidelity control variates), (ii)~concentrate computational effort in rare-event regions using lower-fidelity models to identify important regions (multifidelity importance sampling), or (iii)~identify where and at what fidelity additional evaluations are most informative to reduce uncertainty in approximations for high-fidelity predictions (multifidelity active learning). Multifidelity methods can also exploit temporal or workflow-level information reuse, treating previously generated data, samples, or surrogate models as additional information sources. Box~\ref{box:mf} shows the key concepts for multifidelity methods described in this review.
\begin{mybox}[box:mf]{Key concepts in multifidelity methods}
\renewcommand{\arraystretch}{1.15}
\begin{tabular}{>{\bfseries}p{0.2\linewidth} p{0.75\linewidth}}
\toprule
\textbf{Term} & \textbf{Meaning} \\
\midrule
High-fidelity model &
A computational model that provides a high-accuracy prediction of the QoI, typically at relatively high computational cost. \\

Low-fidelity model &
A computationally cheaper model that provides an approximation of the QoI. Although generally less accurate than a high-fidelity model, it can provide useful information. \\

Correlation & The correlation measures how strongly the model outputs from different fidelity levels vary together as the uncertain inputs change. \\

Multifidelity methods &
Computational methods that combine information from models with different accuracy and computational costs to estimate a high-fidelity QoI statistic more efficiently than could be done by using the high-fidelity model alone. \\

Budget &
The amount of resources available to conduct a computational task, such as estimation of a risk or reliability measure. Typically budget is practically wall clock time or processing time, but is often expressed in terms of the number of executions of the high-fidelity model.  \\
\bottomrule
\end{tabular}

\vspace{2mm}

\textbf{Multifidelity uncertainty quantification techniques}
\vspace{1mm}

\begin{tikzpicture}[
    box/.style={draw=none, fill=white, rounded corners=8pt, text=darkblue, align=center, minimum height=1.7cm, inner xsep=5pt, inner ysep=5pt,}
]
\definecolor{darkblue}{RGB}{17,47,81}

\node[box, text width=4cm] {
    {\bfseries\fontsize{10}{12}\selectfont Control variates}\\[4pt]
    \fontsize{9}{10}\selectfont
    Exploit correlation across fidelities
};

\node[box, text width=4cm, right=0.3cm of current bounding box.east] {
    {\bfseries\fontsize{10}{12}\selectfont Importance sampling}\\[4pt]
    \fontsize{9}{10}\selectfont
    Use cheap models to identify tail regions
};

\node[box, text width=4.75cm, right=0.3cm of current bounding box.east] {
    {\bfseries\fontsize{10}{12}\selectfont Active learning}\\[4pt]
    \fontsize{9}{10}\selectfont
    Improve surrogate by querying where uncertainty matters
};

\end{tikzpicture}
\end{mybox}

\subsection{Multifidelity control variates} \label{s:mfcv}
Control variates reduce the variance of Monte Carlo estimators by exploiting the correlation between the QoI and an auxiliary quantity whose expectation is known or can be estimated~\cite{Bratley1987,owen2013}. In multifidelity control variates approaches,  the auxiliary quantities are obtained from evaluating lower-fidelity models, enabling information from multiple models of varying fidelity to be combined. Because lower-fidelity models are cheaper to evaluate, many more samples can be obtained from them than from the high-fidelity model. In order to provide computational benefit, the lower-fidelity models require strong correlation (i.e., captures similar variations in the QoI across different realizations of the uncertain inputs) and sufficiently lower cost with respect to the high-fidelity model. Stronger correlation between models allows the more frequently evaluated lower-fidelity models to provide information about the high-fidelity response and reduce the variance of the estimated statistics. The multilevel Monte Carlo (MLMC) method uses a control variates approach in which estimators are constructed from a telescoping sum of differences between successive model fidelities~\cite{heinrich2001multilevel,Giles2008MLMC,cliffe2011multilevel,barth2011multi,giles2015multilevelacta,giles2019multilevel}. The multifidelity Monte Carlo (MFMC) method combines the control variates principle with optimal resource allocation across a collection of multifidelity models~\cite{Ng2014MFMCOptimization, Peherstorfer2016MFMC, peherstorfer2019multifidelity, farcaș2023context, gruber2023multifidelity}. The optimal allocation balances the cost of each model against its correlation with the high-fidelity model via an explicit sample allocation strategy that minimizes estimator variance for a given computational budget~\cite{Peherstorfer2016MFMC}. The more general approximate control variate (ACV) estimators~\cite{Gorodetsky2020ACV} and multifidelity best linear unbiased estimator (MLBLUE)~\cite{Schaden2020MLBLUE,schaden2021asymptotic,croci2023multi} relax some of the sampling and nesting assumptions of MFMC and MLMC. 

To introduce the setup and notation, first consider the multifidelity control variate estimator for the expected value, $\mu=\mathbb{E}_{\pi}[y^{(0)}]$: 
\begin{equation}
\mathbb{E}_{\pi}\left[y^{(0)}\right] \approx \frac{1}{N_0} \sum_{i=1}^{N_0} y^{(0)}_i + \sum_{\ell=1}^{L} \beta_\ell \left[ \frac{1}{N_\ell} \sum_{i=1}^{N_\ell} y^{(\ell)}_i - \frac{1}{N_{\ell-1}} \sum_{i=1}^{N_{\ell-1}} y^{(\ell)}_i\right],
\label{eq:mfmc}
\end{equation}
where $y^{(\ell)}_i$ denotes the $i$th QoI sample evaluated using model $\ell$ and $N_\ell$ is the total number of samples of model $\ell$. We have $N_L \geq N_{L-1} \geq \dots \geq N_0 > 0$, that is, cheaper models are sampled more often than more expensive models. 
The first term on the right-hand side of (\ref{eq:mfmc}) is the standard Monte Carlo estimate of the mean using $N_0$ samples of the high-fidelity model $y^{(0)}$.
The second term on the right-hand side of (\ref{eq:mfmc}) provides a correction to this mean estimate using the lower-fidelity models $\ell=1,\ldots,L$, weighted by the control variate coefficients $\beta_\ell, \, \ell=1,\ldots,L$. Notice that the correction terms associated with model $\ell$ is zero in expectation, so that the estimator (\ref{eq:mfmc}) remains unbiased even when the low-fidelity models themselves have error.
The advantage of introducing the control variate terms is that they reduce the variance of the estimator, so that for a given computational budget (\ref{eq:mfmc}) provides a better estimate of $\mathbb{E}_{\pi}[y^{(0)}]$ than would be obtained by sampling only $y^{(0)}$. 

To illustrate the variance reduction achieved with multifidelity control variates, consider the MFMC method. In the MFMC setup, the samples are nested and reused; that is, the term  $\sum_{i=1}^{N_{\ell-1}} y^{(\ell)}_i$ is computed with samples that are a subset of those used to compute the term $\sum_{i=1}^{N_\ell} y^{(\ell)}_i$. For a fixed computational budget, the sample allocation can be optimized by jointly accounting for the correlation between each lower-fidelity model and the high-fidelity model, and the corresponding model evaluation cost. After solving the optimal MFMC allocation problem, the ratio of mean squared errors (MSEs) of MFMC ($\operatorname{MSE}_\text{MFMC}$) compared to standard single-fidelity Monte Carlo using only the high-fidelity model ($\operatorname{MSE}_\text{MC}$) shows the cost--correlation trade-off required for variance reduction as given by~\cite{Peherstorfer2016MFMC}:
\begin{equation}
\label{eq:mfmc_mse_ratio}
\frac{\operatorname{MSE}_\text{MFMC}}{\operatorname{MSE}_\text{MC}} = \left[\sum_{\ell=0}^{L} \sqrt{\frac{c_\ell}{c_0} \left(
\rho_{0,\ell}^2-\rho_{0,\ell+1}^2 \right)}\right]^2,
\end{equation}
where $\rho_{0,\ell}$ is the correlation between the high-fidelity QoI $y^{(0)}$ and the QoI $y^{(\ell)}$ for $\ell=0,\dots, L$. We note that $\rho_{0,0}=1$ and define $\rho_{0,L+1}=0$. Equation~\eqref{eq:mfmc_mse_ratio} illustrates the trade-off between model cost and correlation in determining the computational benefit of MFMC. Stronger correlation between the high- and low-fidelity models generally enables greater variance reduction, whereas the relative computational cost determines how many additional lower-fidelity evaluations can be afforded. The differences between successive squared correlation coefficients also account for the additional information provided by each fidelity level relative to the preceding models. Thus, a lower-fidelity model is most beneficial when it provides substantial additional correlation with the high-fidelity response at a sufficiently low computational cost.

The multifidelity control variate approach applies directly to the risk and reliability measures introduced in Section~\ref{s:risk} whenever the target statistic can be expressed as an expectation. The PoF, written as the expectation of the high-fidelity failure indicator as in Eq.~\eqref{eq:pof}, can be approximated using multifidelity control variates as
\begin{equation}
\mathbb{E}_{\pi}\left[\mathbb{I}\left[y^{(0)}>y_{\mathrm{th}}\right]\right] \approx \frac{1}{N_0} \sum_{i=1}^{N_0} \mathbb{I}\left[y^{(0)}_i>y_{\mathrm{th}}\right] + \sum_{\ell=1}^{L} \beta_\ell \left[ \frac{1}{N_\ell} \sum_{i=1}^{N_\ell} \mathbb{I}\left[y^{(\ell)}_i>y_{\mathrm{th}}\right]  - \frac{1}{N_{\ell-1}} \sum_{i=1}^{N_{\ell-1}} \mathbb{I}\left[y^{(\ell)}_i>y_{\mathrm{th}}\right]\right].
\label{eq:pof_cv}
\end{equation}
Along these lines, MLMC has been extended to PoF estimation by applying control variates to the indicator function in Eq.~\eqref{eq:pof}~\cite{fagerlund2016multilevel,elfverson2016multilevel}. For CVaR estimation, multifidelity control variates can be applied through the expectation in Eq.~\eqref{eq:cvar}. For a fixed $\gamma$, the quantity $\mathbb{E}_{\pi}\left[\left(y^{(0)}- \gamma\right)_+\right]$
can be estimated using lower-fidelity control variates as
\begin{equation}
    \mathbb{E}_{\pi}\left[\left(y^{(0)}- \gamma\right)_+\right] \approx \frac{1}{N_0} \sum_{i=1}^{N_0} \left(y^{(0)}_i- \gamma\right)_+ + \sum_{\ell=1}^{L} \beta_\ell \left[ \frac{1}{N_\ell} \sum_{i=1}^{N_\ell} \left(y^{(\ell)}_i - \gamma\right)_+ - \frac{1}{N_{\ell-1}} \sum_{i=1}^{N_{\ell-1}} \left(y^{(\ell)}_i- \gamma\right)_+ \right],
\end{equation}
after which the scalar minimization over $\gamma$ can be performed as in Eq.~\eqref{eq:cvar}. Note that multifidelity control variates are effective when lower-fidelity models are correlated with respect to the particular statistic being estimated and not just the raw model output QoI. For PoF estimation, this requires correlation between $\mathbb{I} \left[ y^{(\ell)}>y_{\mathrm{th}} \right]$, which depends on how consistently the models classify the same input realization as safe or failed. For CVaR estimation, the correlation is required between $\left(y^{(\ell)}- \gamma\right)_+$, which measures the exceedance beyond a given threshold $\gamma$.

An alternative to directly estimating individual risk measures is to estimate the  cumulative density function (CDF) of the high-fidelity QoI, $F_{y^{(0)}}(t) = \mathbb{P}(y^{(0)}\le t)$, over a range of response thresholds $t$. Once the CDF has been estimated, several risk and reliability measures, such as PoF and quantile-based measures can be obtained directly from the estimated CDF (see Eqs.~\eqref{eq:pof} and \eqref{eq:var}). Multilevel approaches for CDF estimation include estimators based on smoothing the discontinuous indicator function and adaptive selection of the smoothing parameter~\cite{giles2015multilevel,lu2016improved}. CVaR estimation through MLMC-based CDF estimation is shown in \cite{krumscheid2018multilevel, ayoul2023quantifying}. Recent work has also shown the ACV version of estimating $F_{y^{(0)}}$~\cite{han2024approximate}.

To summarize, multifidelity control variates reduce computational cost by exploiting the correlation between the high- and low-fidelity estimators of the target statistic of the QoI while factoring in the relative computational costs of the models. The multifidelity control variate strategy differs from multifidelity importance sampling described in the next section, which instead uses lower-fidelity information to change where high-fidelity samples are generated. Multifidelity control variates are therefore attractive when lower-fidelity models provide accurate global information about the high-fidelity response, whereas importance sampling is advantageous when the estimator is dominated by localized rare-event regions.

\subsection{Multifidelity importance sampling} \label{s:mfis}

Importance sampling is particularly useful for risk and reliability analysis when the regions that contribute most to the estimator are rarely sampled under the nominal input density $\pi(\bs{\xi})$. For example, when estimating a small PoF, most samples from $\pi(\bs{\xi})$ may correspond to the safe regions, providing limited information about the failure probability. The main idea in importance sampling is to shift the sampling distribution toward the important regions, such as the failure regions, so that these regions are sampled more frequently. Instead of drawing samples directly from $\pi$, importance sampling draws samples from a biasing density $q(\bs{\xi})$ that places more probability mass in important regions and corrects for the change of measure using the likelihood ratio $\pi/q$~\cite{Bratley1987,owen2013}. Again to introduce the setup and notation, we start with rewriting the expectation of the QoI under a biasing density $q$ as
\begin{equation}
\mathbb{E}_{\pi}\left[y^{(0)}\right] = \mathbb{E}_{q} \left[y^{(0)} \frac{\pi(\bs{\xi})}{q(\bs{\xi})}\right]
\approx \frac{1}{N} \sum_{i=1}^{N} y^{(0)}_i \frac{\pi(\bs{\xi}_i)}{q(\bs{\xi}_i)},
\qquad \bs{\xi} \sim q,
\label{eq:is_general}
\end{equation}
such that $\operatorname{supp}(\pi)\subseteq\operatorname{supp}(q)$.
The effectiveness of importance sampling therefore depends on constructing a biasing density that concentrates samples in regions important to the target quantity without excluding regions with non-zero probability under $\pi$. Multifidelity importance sampling (MFIS) approaches use lower-fidelity models to construct or guide this biasing density~\cite{Peherstorfer2016MFIS}. {For MFIS approaches, the lower-fidelity models need to provide useful information about where important events occur with regard to the high-fidelity simulations.} 

The lower-fidelity models can be evaluated at lower computational cost and can therefore be used to explore $\Xi$ more extensively than the high-fidelity model. The role of the lower-fidelity model in MFIS is to identify regions of the uncertain input space where high-fidelity evaluations are likely to be more informative.
For reliability analysis, the failure region estimated using model $\ell$ can be written as
$\mathcal{F}^{(\ell)} = \left\{\bs{\xi}\in\Xi:y^{(\ell)}(\bs{\xi})>y_{\mathrm{th}}\right\}, \ell=0,\ldots,L$. The target reliability measure is determined by the high-fidelity failure region $\mathcal{F}^{(0)}$, but the lower-fidelity regions $\mathcal{F}^{(1)},\ldots,\mathcal{F}^{(L)}$ can provide inexpensive approximations of where important samples are likely to occur, and thus provide information to construct a multifidelity biasing density $q_{\mathrm{MF}}$. We write
$q_{\mathrm{MF}}(\bs{\xi}) = \mathcal{Q} \left( \bs{\xi}; y^{(1)},\ldots,y^{(L)} \right)$, where $\mathcal{Q}$ denotes the procedure used to construct the sampling distribution from the available lower-fidelity information to approximate the theoretically optimal biasing density $q^* = \frac{\mathbb{I}\left[y^{(0)} > y_{\mathrm{th}}\right]\pi(\bs{\xi})}{p_f}$. The theoretical optimal biasing density concentrates all sampling effort in the high-fidelity failure region to eliminate non-informative samples; however, one cannot construct it in practice since it requires the knowledge of the PoF in advance. MFIS addresses this challenge by constructing $q_{\mathrm{MF}}$ to approximate $q^*$ using samples from the hierarchy of lower-fidelity models to capture the failure region and then fitting a Gaussian mixture model to the sampled failure points.
The MFIS estimator for the PoF defined in Eq.~\eqref{eq:pof} is given by
\begin{equation}
\widehat{p}_{f,\mathrm{MFIS}} = \mathbb{E}_{q_{\mathrm{MF}}} \left[\mathbb{I}\left[y^{(0)} > y_{\mathrm{th}}\right] \frac{\pi(\bs{\xi})}{q_{\mathrm{MF}}(\bs{\xi})}\right] 
\approx \frac{1}{N} \sum_{i=1}^{N} \mathbb{I}\left[y^{(0)}_i > y_{\mathrm{th}}\right] \frac{\pi(\bs{\xi}_i)}{q_{\mathrm{MF}}(\bs{\xi}_i)},
\qquad \bs{\xi}\sim q_{\mathrm{MF}}.
\label{eq:mfis_pof}
\end{equation}
The high-fidelity model therefore remains responsible for determining whether a sampled input constitutes a failure. Lower-fidelity models instead determine where high-fidelity models are preferentially evaluated. Provided $\operatorname{supp}(\pi)\subseteq\operatorname{supp}(q_{\mathrm{MF}})$ holds, errors in the lower-fidelity models affect the efficiency of the estimator through the quality of $q_{\mathrm{MF}}$, but the PoF estimator $\widehat{p}_{f,\mathrm{MFIS}}$ remains unbiased. Related strategies are also used in delayed acceptance Markov chain Monte Carlo, where cheaper approximations are first used to screen proposed samples, and the expensive high-fidelity model is evaluated only for proposals that pass the initial screening~\cite{christen2005markov,Lykkegaard2023}.

The same change-of-measure principle can be applied to tail-risk measures. In particular, the expectation appearing in the CVaR formulation in Eq.~\eqref{eq:cvar} can be evaluated using importance sampling, with lower-fidelity models used to concentrate high-fidelity evaluations in regions contributing to the upper tail. There are two approaches to estimating CVaR through MFIS. The first is to rewrite CVaR estimation for a continuous distribution of $y^{(0)}$ as $\operatorname{CVaR}_{\alpha}(y^{(0)})= \mathbb{E}_\pi [y^{(0)} \mid y^{(0)} \ge \operatorname{VaR}_{\alpha}(y^{(0)})]= \frac{1}{1-\alpha}\mathbb{E}_{\pi} \left[ y^{(0)} \, \mathbb{I} [y^{(0)} \ge \operatorname{VaR}_{\alpha}(y^{(0)})] \right]$ and then use MFIS to obtain~\cite{heinkenschloss2018conditional,lee2023multifidelity}
\begin{equation}    \widehat{\operatorname{CVaR}}_{\alpha,\mathrm{MFIS}} \left(y^{(0)}\right) = \frac{1}{1-\alpha}\mathbb{E}_{q_{\mathrm{MF}}} \left[ y^{(0)} \, \mathbb{I} [y^{(0)} \ge \operatorname{VaR}_{\alpha}(y^{(0)})] \frac{\pi(\bs{\xi})}{q_{\mathrm{MF}}(\bs{\xi})}\right], \qquad \bs{\xi}\sim q_{\mathrm{MF}}.
    \label{eq:cvar_mfis}
\end{equation}
In Eq.~\eqref{eq:cvar_mfis}, the $\operatorname{VaR}_{\alpha}(y^{(0)})]$ is typically estimated jointly with the importance sampling procedure.
A second approach is to use MFIS for the expectation appearing in the CVaR formulation in Eq.~\eqref{eq:cvar}, with lower-fidelity models used to concentrate high-fidelity evaluations in regions contributing to the upper tail. For a fixed value of $\gamma$,
\begin{equation}
\mathbb{E}_{\pi} \left[(y^{(0)}-\gamma)_+\right] = \mathbb{E}_{q_{\mathrm{MF}}} \left[\left(y^{(0)}- \gamma \right)_+ \frac{\pi(\bs{\xi})} {q_{\mathrm{MF}}(\bs{\xi})} \right], \qquad \bs{\xi}\sim q_{\mathrm{MF}}.
\label{eq:cvar_mfis2}
\end{equation}
Consequently, the CVaR can be estimated by solving the same scalar minimization problem as in Eq.~\eqref{eq:cvar}, but evaluating its expectation under the multifidelity biasing density. Note that Eq.~\eqref{eq:cvar_mfis2} does not require the distribution of $y^{(0)}$ to be continuous.

Several approaches exploit the idea of creating multifidelity biasing densities. Mixed multifidelity importance sampling extends MFIS to multiple lower-fidelity models by using control variates to combine importance sampling estimators from several fidelity-specific $q_\mathrm{MF}$ distributions~\cite{Peherstorfer2017MMFIS}. Another approach fuses multiple MFIS estimators into a weighted ensemble estimator with the weights selected to minimize the variance of the fused estimator~\cite{kramer2019multifidelity}. Other approaches combine importance sampling with multilevel and multifidelity control variates, using importance sampling to concentrate evaluations in important regions while exploiting correlated lower-fidelity information to further reduce estimator variance~\cite{Kebaier2018MLIS,pham2022ensemble,chakroborty2025covariance}. In the context of adaptive importance sampling strategies, multifidelity cross-entropy (MFCE) uses lower-fidelity models to construct and progressively refine a biasing density before introducing more expensive high-fidelity evaluations~\cite{Peherstorfer2018MFCE}. This provides a systematic way of adapting the sampling distribution to rare-event regions without requiring the high-fidelity model to explore the entire input space. Using Eq.~\eqref{eq:cvar_mfis}, MFCE approaches have also been developed for CVaR estimation and risk-averse optimization~\cite{Chaudhuri2020MFCECVaR}. Related multilevel approaches apply the adaptive biasing density construction idea to sequential importance sampling and subset simulation, using a hierarchy of increasingly accurate model discretizations to progressively concentrate samples toward rare-event regions. Multilevel subset simulation constructs a sequence of intermediate failure domains across model resolutions~\cite{ullmann2015multilevel}, while multilevel sequential importance sampling combines sequential importance sampling with adaptive model refinement to estimate the high-fidelity PoF more efficiently~\cite{wagner2020multilevel}.
Similarly, one can apply MFIS and MFCE approaches for bPoF estimation using the importance sampling estimators described in \cite{harajli2015importance,budhiraja2021minimization}.

To summarize, MFIS approaches reduce computational cost through a mechanism distinct from multifidelity control variates. Whereas control variates use correlated lower-fidelity quantities to reduce the variance of an estimator, MFIS uses lower-fidelity information to change where high-fidelity evaluations are performed. MFIS is particularly useful when the target risk or reliability measure is dominated by a small region of the uncertainty space, such as a rare-event or extreme-tail region.

\subsection{Multifidelity active learning} \label{s:mfal}

Active learning is an adaptive approach to refine an approximate surrogate model by sequentially selecting what information to add next for maximum improvement in the surrogate predictions. Rather than evaluating the computational model at a predetermined set of input locations to fit the surrogate model, active learning uses the accumulated information from previous evaluations to determine most valuable additional evaluations. Multifidelity active learning provides a third strategy for reducing the computational cost of risk and reliability analysis by adaptively selecting (i)~where to evaluate the computational model and (ii)~at what fidelity evaluations are most informative. The objective is to reduce uncertainty in a target risk or reliability estimate at minimum cost. For multifidelity active learning, lower-fidelity models are useful when their evaluations provide information that reduces goal-oriented prediction uncertainty in the high-fidelity response or in a quantity derived from it, such as a failure boundary.

An effective multifidelity probabilistic surrogate model that can quantify uncertainty in predictions as well as provide a natural framework for combining information from multiple fidelity models is central to multifidelity active learning. Gaussian processes (GPs) are particularly well suited to this setting because they provide a probabilistic prediction of the model response at each input location, characterized by a normal distribution of that leads to both a prediction mean and a prediction variance~\cite{rasmussen2005GPsforMLbook}. The predictive mean provides the estimated response, while the prediction variance quantifies the uncertainty associated with that response that can be used to identify regions of input space where additional model evaluations are likely to improve the surrogate. The GP framework also naturally allows information from multiple computational models to be combined into a multifidelity GP. Several multifidelity GPs formulations have been developed that include co-Kriging, autoregressive, nonlinear autoregressive, and more flexible covariance-based formulations~\cite{kennedy2000predicting,forrester2007multi,le2014recursive,perdikaris2015,perdikaris2017nonlinear,poloczek2017multi}. Related multi-output and multi-task GP formulations provide alternative ways to model correlations across fidelity levels~\cite{teh2005semiparametric,bonilla2007multi}. In a multifidelity GP, observations from inexpensive lower-fidelity models can reduce predictive uncertainty in the high-fidelity response, provided that the models exhibit a sufficiently informative statistical relationship. For example, if a lower-fidelity model captures the general variation of the high-fidelity response, observations from that model can improve predictions at input locations where high-fidelity evaluations are unavailable. The multifidelity GP learns this relationship from the available model evaluations while accounting for discrepancies between fidelity levels. Beyond GPs, other probabilistic surrogate models can provide improved scalability for high-dimensional inputs and large datasets and greater flexibility for modeling complex dynamical systems. Some such options for probabilistic surrogate models include probabilistic versions of neural networks, such as Bayesian neural networks, deep ensembles, and dropout-based approaches~\cite{blundell2015weight,gal2016dropout,gal2016dropout, lakshminarayanan2017simple}, as well as physics-aware Bayesian reduced-order models~\cite{guo2022bayesian,mcquarrie2025bayesian,cheng2025adaptive, mcquarrie2026active}. Recent work has begun to extend Bayesian neural networks to multifidelity surrogate modeling~\cite{kerleguer2024bayesian}, but substantial opportunities remain for developing scalable, physics-informed multifidelity probabilistic surrogates. 

The role of multifidelity probabilistic surrogate models in active learning is to quantify how observations at one fidelity reduce uncertainty in the high-fidelity response. The predictive uncertainty of the high-fidelity response is used to construct an acquisition function that balances the information gained from an additional evaluation against the computational cost of gaining that information. The construction of the acquisition function depends on the statistics of interest. For estimating the mean, an acquisition function that reduces global error, such as maximizing prediction variance with a cost penalty, can be suitable; however, for reliability analysis reducing global prediction error is not necessarily the appropriate objective and instead one should identify the failure boundary accurately~\cite{moustapha2022active}. This distinction motivates goal-oriented active learning.

In multifidelity active learning, the next evaluation is selected jointly in terms of its input location $\bs{\xi}$ and fidelity level $\ell$. Let $a(\ell, \bs{\xi};\mathcal{D}, c_\ell)$ denote an acquisition function representing the expected utility of evaluating fidelity $\ell$ at input $\bs{\xi}$, given the current multifidelity training data $\mathcal{D} = \bigcup_{\ell=0}^{L} \mathcal{D}_\ell$ and the fidelity-specific computational cost $c_\ell$, where $\mathcal{D}_\ell=\{ (\xi_i,y^{(\ell)}_i) \}_{i=1}^{N_\ell}, \ell=0,\dots,L$ denotes the $N_\ell$ training samples obtained to date from model $\ell$. A general cost-aware acquisition strategy is 
\begin{equation}
(\ell^{\star},\bs{\xi}^{\star}) =
\underset{\ell\in \{0,\ldots,L\}, \bs{\xi}\in\Xi}{\arg\max}
a\left(\ell,\bs{\xi};\mathcal{D},c_\ell\right).
\label{eq:mf_acquisition_general}
\end{equation}
The optimization statement (\ref{eq:mf_acquisition_general}) shows the two decisions inherent in multifidelity active learning: where to evaluate ($\bs{\xi}^{\star}$) and which fidelity to use ($\ell^\star$). A lower-fidelity evaluation is preferred when it is expected to substantially reduce the uncertainty in the target relevant to the target risk or reliability measure at a relatively small computational cost, whereas a high-fidelity evaluation becomes preferable when lower-fidelity information is insufficient.

For reliability analysis, the PoF depends primarily on the location of the failure boundary of the high-fidelity model output, $\Gamma^{(0)} =
\left\{\boldsymbol{\xi}\in\Xi:y^{(0)}(\boldsymbol{\xi})=y_{\mathrm{th}}\right\}$. Consequently, the most valuable model evaluations are those that reduce the uncertainty about whether points near the failure boundary belong to the safe or the failure domain. The goal-oriented acquisition functions developed for the purpose of accurately identifying the failure boundary include probability of feasibility, expected feasibility function (EFF)~\cite{bichon2008efficient}, U-learning~\cite{echard2011ak}, and integrated measures such as the targeted integrated mean squared error~\cite{picheny2010adaptive}. Although these criteria differ in their definitions of utility of information, they concentrate evaluations in regions where uncertainty in the failure classification or failure boundary is high~\cite{moustapha2022active}. 

Multifidelity active learning can exploit lower-fidelity models to identify regions likely to contain the failure boundary and then determine whether a lower- or higher-fidelity evaluation is most informative. This is particularly useful when the failure boundary is localized in a small portion of $\Xi$; a low-fidelity model may accurately classify most of the input domain as clearly safe or clearly failed, leaving only a narrow region near $y^{(0)}=y_{\mathrm{th}}$ that requires high-fidelity evaluations. The method can therefore focus computational effort on the failure boundary region rather than constructing a globally accurate surrogate. Several multifidelity reliability methods can be viewed within this framework. CLoVER~\cite{Marques2018CLOVER,marques2020multifidelity} uses an entropy-based acquisition function to guide the selection of additional evaluations. mfEGRA~\cite{Chaudhuri2021mfEGRA} uses a two-stage process with EFF for location selection and a weighted information-gain-based criterion for fidelity selection. The two-stage process can be extended to any single-fidelity acquisition function to create a multifidelity analog~\cite{chaudhuri2022generalized}. CAMERA combines a similar cost-aware strategy that can utilize existing single-fidelity acquisition functions with importance sampling for PoF estimation~\cite{renganathan2023camera}. Ref.~\cite{dhulipala2022active} expands U-learning acquisition to a multifidelity setting and combines it with subset simulation. These approaches illustrate different strategies for location and fidelity-selection.

Beyond PoF, a similar goal-oriented formulation applies to other risk measures. For a generic risk or reliability quantity $\mathcal{G}[y^{(0)}]$, an ideal acquisition function would select the next model evaluation according to its expected reduction in uncertainty in $\mathcal{G}$,
\begin{equation}
(\ell^{\star},\bs{\xi}^{\star}) =
\underset{\ell\in \{0,\ldots,L\}, \bs{\xi}\in\Xi}{\arg\max}   \frac{1}{c_\ell} \mathbb{E}\left[\mathcal{U}\left(\mathcal{G}[y^{(0)}] \mid \mathcal{D}\right) - \mathcal{U}\left(\mathcal{G}[y^{(0)}] \mid \mathcal{D}\cup (\bs{\xi},y^{(\ell)}) \right)\right],
\label{eq:risk_active_learning}
\end{equation}
where $\mathcal{U}$ denotes an uncertainty measure associated with the target risk or reliability quantity and the expectation is taken over the possible outcome of the new model evaluation. The computational cost of each fidelity is incorporated into the acquisition function as a normalization factor, so that an inexpensive evaluation is favored when it is expected to provide substantial information about the target quantity, while a more expensive evaluation is selected when lower-fidelity evaluations provide insufficient information. For quantities such as CVaR, the relevant region may extend beyond the failure boundary to the portion of the input space contributing to the upper tail of the response. Goal-oriented acquisition functions for such quantities would therefore target uncertainty in the corresponding tail behavior or directly in the risk estimate as presented in the idealized acquisition function in Eq.~\eqref{eq:risk_active_learning}. Developing efficient multifidelity acquisition functions specifically tailored to such risk measures remains an important direction for future work.

To summarize, multifidelity active learning reduces the cost of risk and reliability analysis by adaptively allocating model evaluations to the locations and fidelity levels that provide the greatest information about the target quantity. Within this framework, multifidelity probabilistic surrogate models are a key ingredient because they combine information from multiple sources while quantifying predictive uncertainty. Multifidelity active learning is complementary to multifidelity control variates and multifidelity importance sampling: rather than modifying an estimator or its sampling distribution, active learning reduces the number of high-fidelity evaluations required to construct a sufficiently accurate estimate of the target risk or reliability measure.

\section{Outlook and future perspectives} \label{s:outlook}

\paragraph{Multifidelity optimization under uncertainty.} 
An important step is to integrate multifidelity risk and reliability analysis into decision-making through optimization under uncertainty (OUU) formulations, where uncertainty quantification is performed repeatedly as the design evolves through the optimization iterations. For example, OUU can be used for risk-informed treatment decisions in oncology digital twins as illustrated in Figure~\ref{fig:oncorisk} (panel E). Let $\bs{x}\in\mathcal{X}$ denote the vector of design variables within the design space $\mathcal{X}$ that represent the decisions to be made, such as shape definition, process variables, and control actions. In the OUU setting, computational model $\ell$ now maps both the uncertain inputs $\bs{\xi}$ and the design variables $\bs{x}$ to a QoI $y^{(\ell)}$, which we write as $y^{(\ell)} = f^{(\ell)}(\bs{x},\bs{\xi})$, where $f^{(\ell)}:\mathcal{X}\times\Xi\rightarrow\mathbb{R}$ denotes the model response for design $\bs{x}$ and uncertain input $\bs{\xi}$ as estimated by model $\ell$. 
The goal is to solve the high-fidelity OUU problem, which can be generically written as
\begin{equation}
\bs{x}^{\star} = \argmin_{\bs{x}\in\mathcal{X}} \mathcal{J}^{(0)}(\bs{x})
\quad \text{subject to} \quad
\mathcal{R}^{(0)}(\bs{x})
\leq 0,
\label{eq:mf_ouu}
\end{equation}
where $\mathcal{J}^{(0)}$ and $\mathcal{R}^{(0)}$ are respectively the objective function and constraint function evaluated with the high-fidelity model. Both the objective function and constraint function might include risk or reliability measures such as PoF, bPoF, or CVaR~\cite{rockafellar2000optimization,RTRockafellar_JORoyset_2015a,chaudhuri2022certifiable,rockafellar2013fundamental}. 
The computational approaches to estimate these measures are the same as discussed in Section~\ref{s:mfrisk} (i.e., using sampling to integrate over the uncertain input $\bs{\xi}$) but the OUU setting requires repeated estimation of the risk or reliability measure for each value of the design variables $\bs{x}$ visited during the optimization iterations.    
Multifidelity methods can substantially reduce the computational cost of the repeated uncertainty quantification required at each optimization iteration, particularly for operational or time-sensitive decision-making. Multifidelity control variate approaches have been used for robust optimization with objectives involving linear combinations of mean and variance~\cite{Ng2014MFMCOptimization,chaudhuri2018multifidelity,vanbarel2019,hyun2026robust}, as well as for CVaR-based optimization~\cite{GANESH2023112523}. 
The systematic extension of these multifidelity methods to bPoF-based optimization remains relatively unexplored. More broadly, despite emerging work on multifidelity OUU, fundamental questions remain regarding how to optimally allocate computational effort jointly across the design space, uncertain input space, and model fidelities. The multidimensional resource allocation and problem formulation for multifidelity OUU spans decisions over $\{\mathcal{R}, \bs{x}, \bs{\xi},\ell\}$ to answer: (i) which risk or reliability measure to use in the OUU problem formulation, (ii) where in the design space to evaluate, (iii) how many uncertain input samples are required, (iv) where in the uncertain input space should those samples be concentrated, given the selected risk or reliability measure, and (v) which fidelity to use. An additional challenge arises when model fidelity is input-dependent, such that the correlation between high- and low-fidelity models varies across the design space and changes as the optimizer explores different regions. Finally, beyond multifidelity estimation of risk and reliability themselves, scalable high-dimensional OUU also requires efficient multifidelity estimation of their gradients with respect to the design variables.

\paragraph{Information reuse as a multifidelity information source.}
An alternative way to construct multifidelity methods is to treat information generated earlier in a computational workflow as an additional source of correlated information. Information reuse exploits previously generated model evaluations, surrogate models, or sampling distributions when estimating a risk or reliability measure at a new design point or iteration. This perspective can be implemented through several complementary variance reduction and active learning mechanisms. In a control variate setting, model evaluations obtained at a previous design point can be incorporated as correlated control variates for a new risk or reliability estimate, provided that the responses remain sufficiently correlated~\cite{Ng2016InformationReuseAircraft,cook2018generalized, law2023meta}. For importance sampling, previously constructed biasing densities and simulation data can be reused at nearby design points rather than restarting the sampling procedure, thus avoiding repeated identification of important regions~\cite{beaurepaire2013reliability,Chaudhuri2020IRIS,ehre2021conditional}. Information reuse can also be combined with multifidelity active learning, where surrogates fit during earlier design iterations are retained and used as additional information sources for subsequent risk and reliability estimation in an optimization workflow~\cite{Chaudhuri2019MFGPIR}. Probabilistic re-analysis is a related line of research that can also be interpreted as an information reuse approach~\cite{kuczera2009rbdo,kuczera2010re,chaudhuri2015nasa}. These illustrate a broader opportunity to view information accumulated throughout an evolving computational workflow as a dynamic multifidelity resource, potentially enabling control variates, importance sampling, and active learning to share and reuse information across repeated risk and reliability analyses, specifically in optimization under uncertainty frameworks.

\paragraph{Downstream applications and integration into digital twins.}
Multifidelity risk and reliability analysis can support design, certification, and operational decision-making in safety-critical applications. In aerospace engineering, for example, high-fidelity computational fluid dynamics and structural mechanics models are used to assess the performance under uncertain operating conditions. Repeated reliability and tail-risk evaluations during design optimization and certification can be computationally prohibitive. Multifidelity methods address this challenge by combining detailed simulations with reduced-order models, coarser discretizations, and data-driven surrogates. Similar opportunities arise in energy infrastructure, where digital twins of wind turbines and power systems could use multifidelity methods to continuously assess system reliability as operating conditions and sensor measurements evolve.
In medicine, multifidelity methods can advance patient-specific digital twins for precision medicine. As illustrated in Figure~\ref{fig:oncorisk}, a hierarchy of mechanistic and data-driven models can predict patient-specific treatment outcomes while accounting for uncertainty in model parameters, measurements, and treatment response. Risk and reliability measures can quantify the likelihood of inadequate tumor control or adverse treatment outcomes and support comparisons among candidate treatment strategies. Beyond individual patients, similar approaches could enable population-level risk assessment by characterizing variability in treatment outcomes across patient populations and identifying vulnerable patient subgroups.

Realizing these opportunities requires development of application-specific model hierarchies that preserve information relevant to target risk measures, which needs domain expertise. Future research should therefore develop application-aware multifidelity frameworks that integrate physical knowledge, observational data, and computational models. Such multifidelity frameworks can make risk and reliability assessment a routine component of next-generation digital twins rather than a computationally prohibitive analysis performed only during isolated design or assessment stages.

\section*{Acknowledgments}
The authors acknowledge support from AFOSR grant FA9550-24-1-0327 under the Multidisciplinary University Research Initiative (MURI), NSF FDT-Biotech award 2436499, DARPA TRS award HR00112590114, and Department of Energy MMMIC grant DE-SC002317.

\bibliographystyle{aiaa}
\bibliography{references}

@article{RTRockafellar_SUryasev_2002a,
  title        = {Conditional value-at-risk for general loss distributions},
  author       = {R. T. Rockafellar and S. Uryasev},
  year         = 2002,
  journal      = {Journal of Banking \& Finance},
  volume       = 26,
  number       = 7,
  pages        = {1443--1471},
  doi          = {10.1016/S0378-4266(02)00271-6},
  fauthor      = {Rockafellar, R. Tyrrell and Uryasev, Stan}
}

@article{rockafellar2000optimization,
  title        = {Optimization of conditional value-at-risk},
  author       = {Rockafellar, R Tyrrell and Uryasev, Stanislav},
  year         = 2000,
  journal      = {Journal of Risk},
  volume       = 2,
  pages        = {21--42}
}

@incollection{rockafellar2013superquantiles,
  title        = {Superquantiles and their applications to risk, random variables, and regression},
  author       = {Rockafellar, R Tyrrell and Royset, Johannes O},
  year         = 2013,
  booktitle    = {Theory Driven by Influential Applications},
  publisher    = {Informs},
  pages        = {151--167}
}

@article{RTRockafellar_JORoyset_2015a,
  title        = {Engineering decisions under risk averseness},
  author       = {R. T. Rockafellar and J. O. Royset},
  year         = 2015,
  journal      = {ASCE-ASME Journal of Risk and Uncertainty in Engineering Systems, Part A: Civil Engineering},
  volume       = 1,
  number       = 2,
  pages        = {04015003},
  doi          = {10.1061/AJRUA6.0000816}
}

@article{rockafellar2010buffered,
  title        = {On buffered failure probability in design and optimization of structures},
  author       = {Rockafellar, R Tyrrell and Royset, Johannes O},
  year         = 2010,
  journal      = {Reliability Engineering \& System Safety},
  publisher    = {Elsevier},
  volume       = 95,
  number       = 5,
  pages        = {499--510}
}

@article{mafusalov2018buffered,
  title        = {Buffered probability of exceedance: mathematical properties and optimization},
  author       = {Mafusalov, Alexander and Uryasev, Stan},
  year         = 2018,
  journal      = {SIAM Journal on Optimization},
  publisher    = {SIAM},
  volume       = 28,
  number       = 2,
  pages        = {1077--1103}
}

@article{norton2019calculating,
  title        = {Calculating {CVaR} and {bPOE} for common probability distributions with application to portfolio optimization and density estimation},
  author       = {Norton, Matthew and Khokhlov, Valentyn and Uryasev, Stan},
  year         = 2019,
  journal      = {Annals of Operations Research},
  publisher    = {Springer},
  pages        = {1--35}
}

@inproceedings{basova2011computational,
  title        = {A computational study of the buffered failure probability in reliability-based design optimization},
  author       = {Basova, H. G. and Rockafellar, R. T. and Royset, J. O.},
  year         = 2011,
  booktitle    = {Proceedings of International Conference on Applications of Statistics and Probability in Civil Engineering (ICASP), Zurich, Switzerland},
  publisher    = {Citeseer}
}

@article{chaudhuri2022certifiable,
  title        = {Certifiable risk-based engineering design optimization},
  author       = {Chaudhuri, Anirban and Kramer, Boris and Norton, Matthew and Royset, Johannes O and Willcox, Karen},
  year         = 2022,
  journal      = {AIAA Journal},
  publisher    = {American Institute of Aeronautics and Astronautics},
  volume       = 60,
  number       = 2,
  pages        = {551--565}
}

@book{Bratley1987,
  title        = {A Guide to Simulation},
  author       = {Bratley, Paul and Fox, Bennett L. and Schrage, Linus E.},
  year         = 1987,
  publisher    = {Springer New York},
  address      = {New York, NY},
  pages        = {44--76},
  doi          = {10.1007/978-1-4419-8724-2_2},
  isbn         = {978-1-4419-8724-2},
  url          = {https://doi.org/10.1007/978-1-4419-8724-2_2},
  chapter      = {Variance Reduction}
}

@book{owen2013,
  title        = {Monte Carlo theory, methods and examples},
  author       = {Art B. Owen},
  year         = 2013,
  publisher    = {\url{https://artowen.su.domains/mc/}}
}

@article{Giles2008MLMC,
  title        = {Multilevel Monte Carlo Path Simulation},
  author       = {Giles, Michael B.},
  year         = 2008,
  journal      = {Operations Research},
  volume       = 56,
  number       = 3,
  pages        = {607--617},
  doi          = {10.1287/opre.1070.0496}
}

@article{giles2015multilevelacta,
  title        = {Multilevel {M}onte {C}arlo methods},
  author       = {Giles, Michael B},
  year         = 2015,
  journal      = {Acta Numerica},
  publisher    = {Cambridge University Press},
  volume       = 24,
  pages        = {259--328}
}

@article{barth2011multi,
  title        = {Multi-level Monte Carlo finite element method for elliptic PDEs with stochastic coefficients},
  author       = {Barth, Andrea and Schwab, Christoph and Zollinger, Nathaniel},
  year         = 2011,
  journal      = {Numerische Mathematik},
  publisher    = {Springer},
  volume       = 119,
  number       = 1,
  pages        = {123--161}
}

@article{cliffe2011multilevel,
  title        = {Multilevel Monte Carlo methods and applications to elliptic PDEs with random coefficients},
  author       = {Cliffe, K. A. and Giles, Michael B and Scheichl, R. and Teckentrup, A. L.},
  year         = 2011,
  journal      = {Computing and Visualization in Science},
  publisher    = {Springer},
  volume       = 14,
  number       = 1,
  pages        = 3
}

@article{Peherstorfer2016MFMC,
  title        = {Optimal Model Management for Multifidelity Monte Carlo Estimation},
  author       = {Peherstorfer, Benjamin and Willcox, Karen and Gunzburger, Max},
  year         = 2016,
  journal      = {SIAM Journal on Scientific Computing},
  volume       = 38,
  number       = 5,
  pages        = {A3163--A3194},
  doi          = {10.1137/15M1046472}
}

@article{peherstorfer2019multifidelity,
  title        = {Multifidelity Monte Carlo estimation with adaptive low-fidelity models},
  author       = {Peherstorfer, Benjamin},
  year         = 2019,
  journal      = {SIAM/ASA Journal on Uncertainty Quantification},
  publisher    = {SIAM},
  volume       = 7,
  number       = 2,
  pages        = {579--603}
}

@article{Ng2014MFMCOptimization,
  title        = {Multifidelity Approaches for Optimization Under Uncertainty},
  author       = {Ng, Leo W. T. and Willcox, Karen E.},
  year         = 2014,
  journal      = {International Journal for Numerical Methods in Engineering},
  volume       = 100,
  number       = 10,
  pages        = {746--772},
  doi          = {10.1002/nme.4761}
}

@article{Ng2016InformationReuseAircraft,
  title        = {Monte Carlo Information-Reuse Approach to Aircraft Conceptual Design Optimization Under Uncertainty},
  author       = {Ng, Leo W. T. and Willcox, Karen E.},
  year         = 2016,
  journal      = {Journal of Aircraft},
  volume       = 53,
  number       = 2,
  pages        = {427--438},
  doi          = {10.2514/1.C033352}
}

@article{gruber2023multifidelity,
  title        = {A multifidelity Monte Carlo method for realistic computational budgets},
  author       = {Gruber, Anthony and Gunzburger, Max and Ju, Lili and Wang, Zhu},
  year         = 2023,
  journal      = {Journal of Scientific Computing},
  publisher    = {Springer},
  volume       = 94,
  number       = 1,
  pages        = 2
}

@article{farcaș2023context,
  title        = {Context-aware learning of hierarchies of low-fidelity models for multi-fidelity uncertainty quantification},
  author       = {Farcaș, Ionuț-Gabriel and Peherstorfer, Benjamin and Neckel, Tobias and Jenko, Frank and Bungartz, Hans-Joachim},
  year         = 2023,
  journal      = {Computer Methods in Applied Mechanics and Engineering},
  publisher    = {Elsevier},
  volume       = 406,
  pages        = 115908
}

@article{Schaden2020MLBLUE,
  title        = {On Multilevel Best Linear Unbiased Estimators},
  author       = {Schaden, Daniel and Ullmann, Elisabeth},
  year         = 2020,
  journal      = {SIAM/ASA Journal on Uncertainty Quantification},
  volume       = 8,
  number       = 2,
  pages        = {801--838},
  doi          = {10.1137/19M1263534}
}

@article{schaden2021asymptotic,
  title        = {Asymptotic analysis of multilevel best linear unbiased estimators},
  author       = {Schaden, Daniel and Ullmann, Elisabeth},
  year         = 2021,
  journal      = {SIAM/ASA Journal on Uncertainty Quantification},
  publisher    = {SIAM},
  volume       = 9,
  number       = 3,
  pages        = {953--978}
}

@article{croci2023multi,
  title        = {Multi-output multilevel best linear unbiased estimators via semidefinite programming},
  author       = {Croci, Matteo and Willcox, Karen E and Wright, Stephen J},
  year         = 2023,
  journal      = {Computer Methods in Applied Mechanics and Engineering},
  publisher    = {Elsevier},
  volume       = 413,
  pages        = 116130
}

@article{Gorodetsky2020ACV,
  title        = {A Generalized Approximate Control Variate Framework for Multifidelity Uncertainty Quantification},
  author       = {Gorodetsky, Alex A. and Geraci, Gianluca and Eldred, Michael S. and Jakeman, John D.},
  year         = 2020,
  journal      = {Journal of Computational Physics},
  volume       = 408,
  pages        = 109257,
  doi          = {10.1016/j.jcp.2020.109257}
}

@article{christen2005markov,
  title        = {Markov chain Monte Carlo using an approximation},
  author       = {Christen, J Andr{\'e}s and Fox, Colin},
  year         = 2005,
  journal      = {Journal of Computational and Graphical Statistics},
  publisher    = {Taylor \& Francis},
  volume       = 14,
  number       = 4,
  pages        = {795--810}
}

@article{Lykkegaard2023,
  title        = {Multilevel Delayed Acceptance MCMC},
  author       = {Lykkegaard, M. B. and Dodwell, T. J. and Fox, C. and Mingas, G. and Scheichl, R.},
  year         = 2023,
  journal      = {SIAM/ASA Journal on Uncertainty Quantification},
  volume       = 11,
  number       = 1,
  pages        = {1--30},
  doi          = {10.1137/22M1476770}
}

@article{Peherstorfer2016MFIS,
  title        = {Multifidelity importance sampling},
  author       = {Peherstorfer, Benjamin and Cui, Tiangang and Marzouk, Youssef and Willcox, Karen},
  year         = 2016,
  journal      = {Computer Methods in Applied Mechanics and Engineering},
  volume       = 300,
  pages        = {490--509},
  doi          = {10.1016/j.cma.2015.12.002}
}

@article{Peherstorfer2017MMFIS,
  title        = {Combining multiple surrogate models to accelerate failure probability estimation with expensive high-fidelity models},
  author       = {Peherstorfer, Benjamin and Kramer, Boris and Willcox, Karen},
  year         = 2017,
  journal      = {Journal of Computational Physics},
  volume       = 341,
  pages        = {61--75},
  doi          = {10.1016/j.jcp.2017.04.012}
}

@article{kramer2019multifidelity,
  title        = {Multifidelity probability estimation via fusion of estimators},
  author       = {Kramer, Boris and Marques, Alexandre Noll and Peherstorfer, Benjamin and Villa, Umberto and Willcox, Karen},
  year         = 2019,
  journal      = {Journal of Computational Physics},
  publisher    = {Elsevier},
  volume       = 392,
  pages        = {385--402}
}

@article{Peherstorfer2018MFCE,
  title        = {Multifidelity preconditioning of the cross-entropy method for rare event simulation and failure probability estimation},
  author       = {Peherstorfer, Benjamin and Kramer, Boris and Willcox, Karen},
  year         = 2018,
  journal      = {SIAM/ASA Journal on Uncertainty Quantification},
  volume       = 6,
  number       = 2,
  pages        = {737--761},
  doi          = {10.1137/17M1122992}
}

@inproceedings{Chaudhuri2020MFCECVaR,
  title        = {Multifidelity Cross-Entropy Estimation of Conditional Value-at-Risk for Risk-Averse Design Optimization},
  author       = {Chaudhuri, Anirban and Peherstorfer, Benjamin and Willcox, Karen},
  year         = 2020,
  booktitle    = {AIAA SciTech 2020 Forum},
  publisher    = {AIAA},
  doi          = {10.2514/6.2020-2129}
}

@article{Chaudhuri2020IRIS,
  title        = {Information Reuse for Importance Sampling in Reliability-Based Design Optimization},
  author       = {Chaudhuri, Anirban and Kramer, Boris and Willcox, Karen},
  year         = 2020,
  journal      = {Reliability Engineering \& System Safety},
  volume       = 201,
  pages        = 106853,
  doi          = {10.1016/j.ress.2020.106853}
}

@inproceedings{Marques2018CLOVER,
  title        = {Contour Location via Entropy Reduction Leveraging Multiple Information Sources},
  author       = {Marques, Alexandre N. and Lam, Remi R. and Willcox, Karen E.},
  year         = 2018,
  booktitle    = {Advances in Neural Information Processing Systems},
  volume       = 31,
  pages        = {5217--5227}
}

@article{marques2020multifidelity,
  title        = {Multifidelity method for locating aeroelastic flutter boundaries},
  author       = {Marques, Alexandre N and Opgenoord, Max MJ and Lam, Remi R and Chaudhuri, Anirban and Willcox, Karen E},
  year         = 2020,
  journal      = {AIAA Journal},
  publisher    = {American Institute of Aeronautics and Astronautics},
  volume       = 58,
  number       = 4,
  pages        = {1772--1784}
}

@article{Chaudhuri2021mfEGRA,
  title        = {mfEGRA: Multifidelity Efficient Global Reliability Analysis through Active Learning for Failure Boundary Location},
  author       = {Chaudhuri, Anirban and Marques, Alexandre N. and Willcox, Karen E.},
  year         = 2021,
  journal      = {Structural and Multidisciplinary Optimization},
  volume       = 64,
  number       = 2,
  pages        = {797--811},
  doi          = {10.1007/s00158-021-02892-5}
}

@inproceedings{Chaudhuri2019MFGPIR,
  title        = {Reusing Information for Multifidelity Active Learning in Reliability-Based Design Optimization},
  author       = {Chaudhuri, Anirban and Marques, Alexandre N. and Lam, Remi and Willcox, Karen E.},
  year         = 2019,
  booktitle    = {AIAA SciTech 2019 Forum},
  doi          = {10.2514/6.2019-1222}
}

@article{dhulipala2022active,
  title        = {Active learning with multifidelity modeling for efficient rare event simulation},
  author       = {Dhulipala, Somayajulu LN and Shields, Michael D and Spencer, Benjamin W and Bolisetti, Chandrakanth and Slaughter, Andrew E and Labour{\'e}, Vincent M and Chakroborty, Promit},
  year         = 2022,
  journal      = {Journal of Computational Physics},
  publisher    = {Elsevier},
  volume       = 468,
  pages        = 111506
}

@article{renganathan2023camera,
  title        = {CAMERA: A method for cost-aware, adaptive, multifidelity, efficient reliability analysis},
  author       = {Renganathan, S Ashwin and Rao, Vishwas and Navon, Ionel M},
  year         = 2023,
  journal      = {Journal of Computational Physics},
  publisher    = {Elsevier},
  volume       = 472,
  pages        = 111698
}

@article{bichon2008efficient,
  title        = {Efficient global reliability analysis for nonlinear implicit performance functions},
  author       = {Bichon, Barron J and Eldred, Michael S and Swiler, Laura Painton and Mahadevan, Sandaran and McFarland, John M},
  year         = 2008,
  journal      = {AIAA Journal},
  volume       = 46,
  number       = 10,
  pages        = {2459--2468}
}

@article{echard2011ak,
  title        = {AK-MCS: an active learning reliability method combining Kriging and Monte Carlo simulation},
  author       = {Echard, Benjamin and Gayton, Nicolas and Lemaire, Maurice},
  year         = 2011,
  journal      = {Structural Safety},
  publisher    = {Elsevier},
  volume       = 33,
  number       = 2,
  pages        = {145--154}
}

@article{picheny2010adaptive,
  title        = {Adaptive Designs of Experiments for Accurate Approximation of a Target Region},
  author       = {Picheny, Victor and Ginsbourger, David and Roustant, Olivier and Haftka, Raphael T. and Kim, Nam-Ho},
  year         = 2010,
  month        = {06},
  journal      = {Journal of Mechanical Design},
  volume       = 132,
  number       = 7,
  pages        = {071008},
  doi          = {10.1115/1.4001873},
  issn         = {1050-0472},
  url          = {https://doi.org/10.1115/1.4001873},
  eprint       = {https://asmedigitalcollection.asme.org/mechanicaldesign/article-pdf/132/7/071008/5807381/071008_1.pdf}
}

@misc{harajli2015importance,
  title        = {Importance sampling in the evaluation and optimization of buffered failure probability},
  author       = {Harajli, Marwan M and Rockafellar, R Tyrrell and Royset, Johannes O},
  year         = 2015
}

@article{budhiraja2021minimization,
  title        = {Minimization of a class of rare event probabilities and buffered probabilities of exceedance},
  author       = {Budhiraja, Amarjit and Lu, Shu and Yu, Yang and Tran-Dinh, Quoc},
  year         = 2021,
  journal      = {Annals of Operations Research},
  publisher    = {Springer},
  volume       = 302,
  number       = 1,
  pages        = {49--83}
}

@article{giles2019multilevel,
  title        = {Multilevel nested simulation for efficient risk estimation},
  author       = {Giles, Michael B and Haji-Ali, Abdul-Lateef},
  year         = 2019,
  journal      = {SIAM/ASA Journal on Uncertainty Quantification},
  publisher    = {SIAM},
  volume       = 7,
  number       = 2,
  pages        = {497--525}
}

@inproceedings{heinrich2001multilevel,
  title        = {Multilevel {M}onte {C}arlo methods},
  author       = {Heinrich, Stefan},
  year         = 2001,
  booktitle    = {International Conference on Large-Scale Scientific Computing},
  pages        = {58--67},
  organization = {Springer}
}

@book{rasmussen2005GPsforMLbook,
  title        = {Gaussian Processes for Machine Learning},
  author       = {Rasmussen, Carl Edward and Williams, Christopher K. I.},
  year         = 2005,
  month        = 11,
  publisher    = {The MIT Press},
  doi          = {10.7551/mitpress/3206.001.0001},
  isbn         = 9780262256834
}

@article{kennedy2000predicting,
  title        = {Predicting the output from a complex computer code when fast approximations are available},
  author       = {Kennedy, Marc C and O'Hagan, Anthony},
  year         = 2000,
  journal      = {Biometrika},
  publisher    = {Oxford University Press},
  volume       = 87,
  number       = 1,
  pages        = {1--13},
  doi          = {10.1093/biomet/87.1.1}
}

@article{le2014recursive,
  title        = {Recursive co-kriging model for design of computer experiments with multiple levels of fidelity},
  author       = {Le Gratiet, Loic and Garnier, Josselin},
  year         = 2014,
  journal      = {International Journal for Uncertainty Quantification},
  publisher    = {Begel House Inc.},
  volume       = 4,
  number       = 5,
  pages        = {365--386},
  doi          = {10.1615/Int.J.UncertaintyQuantification.2014006914}
}

@article{perdikaris2015,
    author = {Perdikaris, P. and Venturi, D. and Royset, J. O. and Karniadakis, G. E.},
    title = {Multi-fidelity modelling via recursive co-kriging and Gaussian–Markov random fields},
    journal = {Proceedings of the Royal Society A: Mathematical, Physical and Engineering Sciences},
    volume = {471},
    number = {2179},
    pages = {20150018},
    year = {2015},
    month = {07},
    issn = {1364-5021},
    doi = {10.1098/rspa.2015.0018}
}

@article{perdikaris2017nonlinear,
  title        = {Nonlinear information fusion algorithms for data-efficient multi-fidelity modelling},
  author       = {Perdikaris, Paris and Raissi, Maziar and Damianou, Andreas and Lawrence, Neil D. and Karniadakis, George Em},
  year         = 2017,
  journal      = {Proceedings of the Royal Society A: Mathematical, Physical and Engineering Sciences},
  volume       = 473,
  number       = 2198,
  pages        = 20160751,
  doi          = {10.1098/rspa.2016.0751},
  issn         = {1364-5021}
}

@article{forrester2007multi,
  title        = {Multi-fidelity optimization via surrogate modelling},
  author       = {Forrester, Alexander IJ and S{\'o}bester, Andr{\'a}s and Keane, Andy J},
  year         = 2007,
  journal      = {Proceedings: Mathematical, Physical and Engineering Sciences},
  publisher    = {JSTOR},
  pages        = {3251--3269}
}

@inproceedings{poloczek2017multi,
  title        = {Multi-information source optimization},
  author       = {Poloczek, Matthias and Wang, Jialei and Frazier, Peter I.},
  year         = 2017,
  booktitle    = {Advances in Neural Information Processing Systems},
  volume       = 30
}

@article{bonilla2007multi,
  title        = {Multi-task Gaussian process prediction},
  author       = {Bonilla, Edwin V and Chai, Kian and Williams, Christopher},
  year         = 2007,
  journal      = {Advances in neural information processing systems},
  volume       = 20
}

@inproceedings{teh2005semiparametric,
  title        = {Semiparametric latent factor models},
  author       = {Teh, Yee Whye and Seeger, Matthias and Jordan, Michael I},
  year         = 2005,
  booktitle    = {International Workshop on Artificial Intelligence and Statistics},
  pages        = {333--340},
  organization = {PMLR}
}

@article{moustapha2022active,
  title        = {Active learning for structural reliability: Survey, general framework and benchmark},
  author       = {Moustapha, Maliki and Marelli, Stefano and Sudret, Bruno},
  year         = 2022,
  journal      = {Structural Safety},
  publisher    = {Elsevier},
  volume       = 96,
  pages        = 102174
}

@inproceedings{chaudhuri2022generalized,
  title        = {Generalized multifidelity active learning for {G}aussian-process-based reliability analysis},
  author       = {Chaudhuri, Anirban and Willcox, Karen},
  year         = 2022,
  booktitle    = {Dynamic Data Driven Applications Systems},
  publisher    = {Springer Nature Switzerland},
  pages        = {17--26}
}

@article{lee2023multifidelity,
  title        = {Multifidelity conditional value-at-risk estimation by dimensionally decomposed generalized polynomial chaos-Kriging},
  author       = {Lee, Dongjin and Kramer, Boris},
  year         = 2023,
  journal      = {Reliability Engineering \& System Safety},
  publisher    = {Elsevier},
  volume       = 235,
  pages        = 109208
}

@article{heinkenschloss2018conditional,
  title        = {Conditional-value-at-risk estimation via reduced-order models},
  author       = {Heinkenschloss, Matthias and Kramer, Boris and Takhtaganov, Timur and Willcox, Karen},
  year         = 2018,
  journal      = {SIAM/ASA Journal on Uncertainty Quantification},
  publisher    = {SIAM},
  volume       = 6,
  number       = 4,
  pages        = {1395--1423}
}

@article{fagerlund2016multilevel,
  title        = {Multilevel {M}onte {C}arlo methods for computing failure probability of porous media flow systems},
  author       = {Fagerlund, Fritjof and Hellman, Fredrik and M{\aa}lqvist, Axel and Niemi, Auli},
  year         = 2016,
  journal      = {Advances in Water Resources},
  publisher    = {Elsevier},
  volume       = 94,
  pages        = {498--509}
}

@article{elfverson2016multilevel,
  title        = {A multilevel Monte Carlo method for computing failure probabilities},
  author       = {Elfverson, Daniel and Hellman, Fredrik and M{\aa}lqvist, Axel},
  year         = 2016,
  journal      = {SIAM/ASA Journal on Uncertainty Quantification},
  publisher    = {SIAM},
  volume       = 4,
  number       = 1,
  pages        = {312--330}
}

@article{ullmann2015multilevel,
  title        = {Multilevel estimation of rare events},
  author       = {Ullmann, Elisabeth and Papaioannou, Iason},
  year         = 2015,
  journal      = {SIAM/ASA Journal on Uncertainty Quantification},
  publisher    = {SIAM},
  volume       = 3,
  number       = 1,
  pages        = {922--953}
}

@article{wagner2020multilevel,
  title        = {Multilevel sequential importance sampling for rare event estimation},
  author       = {Wagner, Fabian and Latz, Jonas and Papaioannou, Iason and Ullmann, Elisabeth},
  year         = 2020,
  journal      = {SIAM Journal on Scientific Computing},
  publisher    = {SIAM},
  volume       = 42,
  number       = 4,
  pages        = {A2062--A2087}
}

@article{chakroborty2025covariance,
  title        = {Covariance-Free Bifidelity Control Variates Importance Sampling for Rare Event Reliability Analysis},
  author       = {Chakroborty, Promit and Dhulipala, Somayajulu LN and Shields, Michael D},
  year         = 2025,
  journal      = {SIAM/ASA Journal on Uncertainty Quantification},
  publisher    = {SIAM},
  volume       = 13,
  number       = 3,
  pages        = {1406--1451}
}

@article{pham2022ensemble,
  title        = {Ensemble approximate control variate estimators: Applications to multifidelity importance sampling},
  author       = {Pham, Trung and Gorodetsky, Alex A},
  year         = 2022,
  journal      = {SIAM/ASA Journal on Uncertainty Quantification},
  publisher    = {SIAM},
  volume       = 10,
  number       = 3,
  pages        = {1250--1292}
}

@article{Kebaier2018MLIS,
  title        = {Coupling importance sampling and multilevel Monte Carlo using sample average approximation},
  author       = {Kebaier, A. and Lelong, J.},
  year         = 2018,
  journal      = {Methodology and Computing in Applied Probability},
  volume       = 20,
  number       = 2,
  pages        = {611--641}
}

@article{giles2015multilevel,
  title        = {Multilevel Monte Carlo approximation of distribution functions and densities},
  author       = {Giles, Michael B and Nagapetyan, Tigran and Ritter, Klaus},
  year         = 2015,
  journal      = {SIAM/ASA Journal on Uncertainty Quantification},
  publisher    = {SIAM},
  volume       = 3,
  number       = 1,
  pages        = {267--295}
}

@article{lu2016improved,
  title        = {An improved multilevel Monte Carlo method for estimating probability distribution functions in stochastic oil reservoir simulations},
  author       = {Lu, Dan and Zhang, Guannan and Webster, Clayton and Barbier, Charlotte},
  year         = 2016,
  journal      = {Water Resources Research},
  publisher    = {Wiley Online Library},
  volume       = 52,
  number       = 12,
  pages        = {9642--9660}
}

@article{han2024approximate,
  title        = {An approximate control variates approach to multifidelity distribution estimation},
  author       = {Han, Ruijian and Kramer, Boris and Lee, Dongjin and Narayan, Akil and Xu, Yiming},
  year         = 2024,
  journal      = {SIAM/ASA Journal on Uncertainty Quantification},
  publisher    = {SIAM},
  volume       = 12,
  number       = 4,
  pages        = {1349--1388}
}

@article{krumscheid2018multilevel,
  title        = {Multilevel Monte Carlo approximation of functions},
  author       = {Krumscheid, Sebastian and Nobile, Fabio},
  year         = 2018,
  journal      = {SIAM/ASA Journal on Uncertainty Quantification},
  publisher    = {SIAM},
  volume       = 6,
  number       = 3,
  pages        = {1256--1293}
}

@article{ayoul2023quantifying,
  title        = {Quantifying uncertain system outputs via the Multi-level Monte Carlo Method- Distribution and robustness measures},
  author       = {Ayoul-Guilmard, Quentin and Ganesh, Sundar and Krumscheid, Sebastian and Nobile, Fabio},
  year         = 2023,
  journal      = {International Journal for Uncertainty Quantification},
  publisher    = {Begel House Inc.},
  volume       = 13,
  number       = 5
}

@inproceedings{gal2016dropout,
  title        = {Dropout as a bayesian approximation: Representing model uncertainty in deep learning},
  author       = {Gal, Yarin and Ghahramani, Zoubin},
  year         = 2016,
  booktitle    = {International Conference on Machine Learning},
  pages        = {1050--1059},
  organization = {PMLR}
}

@inproceedings{blundell2015weight,
  title        = {Weight uncertainty in neural network},
  author       = {Blundell, Charles and Cornebise, Julien and Kavukcuoglu, Koray and Wierstra, Daan},
  year         = 2015,
  booktitle    = {International conference on machine learning},
  pages        = {1613--1622},
  organization = {PMLR}
}

@article{lakshminarayanan2017simple,
  title        = {Simple and scalable predictive uncertainty estimation using deep ensembles},
  author       = {Lakshminarayanan, Balaji and Pritzel, Alexander and Blundell, Charles},
  year         = 2017,
  journal      = {Advances in Neural Information Processing Systems},
  volume       = 30
}

@article{kerleguer2024bayesian,
  title        = {A Bayesian neural network approach to multi-fidelity surrogate modeling},
  author       = {Kerleguer, Baptiste and Cannamela, Claire and Garnier, Josselin},
  year         = 2024,
  journal      = {International Journal for Uncertainty Quantification},
  publisher    = {Begel House Inc.},
  volume       = 14,
  number       = 1
}

@article{mcquarrie2025bayesian,
  title        = {Bayesian learning with Gaussian processes for low-dimensional representations of time-dependent nonlinear systems},
  author       = {McQuarrie, Shane A and Chaudhuri, Anirban and Willcox, Karen E and Guo, Mengwu},
  year         = 2025,
  journal      = {Physica D: Nonlinear Phenomena},
  publisher    = {Elsevier},
  volume       = 475,
  pages        = 134572
}

@article{guo2022bayesian,
  title        = {Bayesian operator inference for data-driven reduced-order modeling},
  author       = {Guo, Mengwu and McQuarrie, Shane A and Willcox, Karen E},
  year         = 2022,
  journal      = {Computer Methods in Applied Mechanics and Engineering},
  publisher    = {Elsevier},
  volume       = 402,
  pages        = 115336
}

@article{cheng2025adaptive,
  title        = {Adaptive data-driven probabilistic reduced-order models for parameterized dynamical systems},
  author       = {Cheng, Kai and Zimmermann, Ralf},
  year         = 2025,
  journal      = {SIAM Journal on Scientific Computing},
  publisher    = {SIAM},
  volume       = 47,
  number       = 2,
  pages        = {C426--C450}
}

@article{mcquarrie2026active,
  title={Active learning for data-driven reduced models of parametric differential systems with Bayesian operator inference},
  author={McQuarrie, Shane A and Guo, Mengwu and Chaudhuri, Anirban},
  journal={Journal of Scientific Computing},
  volume={109},
  number={2},
  pages={42},
  year={2026},
  publisher={Springer}
}

@article{cook2018generalized,
  title        = {Generalized information reuse for optimization under uncertainty with non-sample average estimators},
  author       = {Cook, Laurence W and Jarrett, Jerome P and Willcox, Karen E},
  year         = 2018,
  journal      = {International Journal for Numerical Methods in Engineering},
  publisher    = {Wiley Online Library},
  volume       = 115,
  number       = 12,
  pages        = {1457--1476}
}

@article{beaurepaire2013reliability,
  title        = {Reliability-based optimization using bridge importance sampling},
  author       = {Beaurepaire, P and Jensen, HA and Schu{\"e}ller, GI and Valdebenito, MA},
  year         = 2013,
  journal      = {Probabilistic Engineering Mechanics},
  publisher    = {Elsevier},
  volume       = 34,
  pages        = {48--57}
}

@article{ehre2021conditional,
  title        = {Conditional reliability analysis in high dimensions based on controlled mixture importance sampling and information reuse},
  author       = {Ehre, Max and Papaioannou, Iason and Willcox, Karen E and Straub, Daniel},
  year         = 2021,
  journal      = {Computer Methods in Applied Mechanics and Engineering},
  publisher    = {Elsevier},
  volume       = 381,
  pages        = 113826
}

@article{law2023meta,
  title        = {Meta variance reduction for Monte Carlo estimation of energetic particle confinement during stellarator optimization},
  author       = {Law, Frederick and Cerfon, Antoine and Peherstorfer, Benjamin and Wechsung, Florian},
  year         = 2023,
  journal      = {Journal of Computational Physics},
  publisher    = {Elsevier},
  volume       = 495,
  pages        = 112524
}

@article{kuczera2009rbdo,
  title        = {An RBDO Method for Multiple Failure Region Problems using Probabilistic Reanalysis and Approximate Metamodels},
  author       = {Kuczera, Ramon C and Mourelatos, Zissimos P and Latcha, Michael and Nikolaidis, Efstratios},
  year         = 2009,
  journal      = {SAE International Journal of Materials and Manufacturing},
  publisher    = {SAE International},
  volume       = 2,
  number       = 1,
  pages        = {108--120}
}

@article{kuczera2010re,
  title        = {A re-analysis methodology for system RBDO using a trust region approach with local metamodels},
  author       = {Kuczera, Ramon and Mourelatos, Zissimos and Nikolaidis, Efstratios},
  year         = 2010,
  journal      = {SAE International Journal of Materials and Manufacturing},
  publisher    = {SAE International},
  volume       = 3,
  number       = 1,
  pages        = {356--371}
}

@article{chaudhuri2015nasa,
  title        = {NASA uncertainty quantification challenge: an optimization-based methodology and validation},
  author       = {Chaudhuri, Anirban and Waycaster, Garrett and Price, Nathaniel and Matsumura, Taiki and Haftka, Raphael T},
  year         = 2015,
  journal      = {Journal of Aerospace Information Systems},
  publisher    = {American Institute of Aeronautics and Astronautics},
  volume       = 12,
  number       = 1,
  pages        = {10--34}
}

@inproceedings{glaessgen2012digital,
  title        = {The Digital Twin Paradigm for Future {NASA} and {U.S.} Air Force Vehicles},
  author       = {Glaessgen, Edward and Stargel, David},
  year         = 2012,
  booktitle    = {53rd AIAA/ASME/ASCE/AHS/ASC Structures, Structural Dynamics and Materials Conference},
  number       = {AIAA 2012-1818},
  pages        = {1--14},
  doi          = {10.2514/6.2012-1818}
}

@article{tuegel2011digitaltwin,
  title        = {Reengineering Aircraft Structural Life Prediction Using a Digital Twin},
  author       = {Tuegel, Eric J. and Ingraffea, Anthony R. and Eason, Thomas G. and Spottswood, S. Michael},
  year         = 2011,
  journal      = {International Journal of Aerospace Engineering},
  volume       = 2011,
  pages        = 154798,
  doi          = {10.1155/2011/154798}
}

@article{thelen2022review,
  title        = {A Comprehensive Review of Digital Twin -- Part 1: Modeling and Twinning Enabling Technologies},
  author       = {Thelen, Adam and Zhang, Xiaoge and Fink, Olga and Lu, Yan and Ghosh, Sayan and Youn, Byeng D. and Todd, Michael D. and Mahadevan, Sankaran and Hu, Chao and Hu, Zhen},
  year         = 2022,
  journal      = {Structural and Multidisciplinary Optimization},
  volume       = 65,
  number       = 12,
  pages        = 354,
  doi          = {10.1007/s00158-022-03425-4}
}

@book{nasem2023digitaltwins,
  title        = {Foundational Research Gaps and Future Directions for Digital Twins},
  author       = {{National Academies of Sciences, Engineering, and Medicine}},
  year         = 2024,
  publisher    = {The National Academies Press},
  address      = {Washington, DC},
  doi          = {10.17226/26894}
}

@article{ferrari2024mechanicalaerospace,
  title        = {Digital Twins in Mechanical and Aerospace Engineering},
  author       = {Ferrari, Alberto and Willcox, Karen},
  year         = 2024,
  journal      = {Nature Computational Science},
  volume       = 4,
  number       = 3,
  pages        = {178--183},
  doi          = {10.1038/s43588-024-00613-8}
}

@article{kapteyn2021pgm,
  title        = {A Probabilistic Graphical Model Foundation for Enabling Predictive Digital Twins at Scale},
  author       = {Kapteyn, Matthew G. and Pretorius, Jacobus V. R. and Willcox, Karen E.},
  year         = 2021,
  journal      = {Nature Computational Science},
  volume       = 1,
  number       = 5,
  pages        = {337--347},
  doi          = {10.1038/s43588-021-00069-0}
}

@article{rockafellar2013fundamental,
  title        = {The fundamental risk quadrangle in risk management, optimization and statistical estimation},
  author       = {Rockafellar, R Tyrrell and Uryasev, Stan},
  year         = 2013,
  journal      = {Surveys in Operations Research and Management Science},
  publisher    = {Elsevier},
  volume       = 18,
  number       = {1-2},
  pages        = {33--53}
}

@article{hyun2026robust,
  title        = {Robust topology optimization of continuum structures under material uncertainties using multifidelity Monte Carlo},
  author       = {Hyun, Jaeyub and Chaudhuri, Anirban and Willcox, Karen E and Kim, H Alicia},
  year         = 2026,
  journal      = {Computer Methods in Applied Mechanics and Engineering},
  publisher    = {Elsevier},
  volume       = 451,
  pages        = 118714
}

@inproceedings{chaudhuri2018multifidelity,
  title        = {Multifidelity optimization under uncertainty for a tailless aircraft},
  author       = {Chaudhuri, Anirban and Jasa, John and Martins, Joaquim RRA and Willcox, Karen E},
  year         = 2018,
  booktitle    = {2018 AIAA Non-Deterministic Approaches Conference},
  doi          = {10.2514/6.2018-1658}
}

@article{vanbarel2019,
  title        = {Robust Optimization of PDEs with Random Coefficients Using a Multilevel Monte Carlo Method},
  author       = {Van Barel, Andreas and Vandewalle, Stefan},
  year         = 2019,
  journal      = {SIAM/ASA Journal on Uncertainty Quantification},
  volume       = 7,
  number       = 1,
  pages        = {174--202},
  doi          = {10.1137/17M1155892}
}

@article{GANESH2023112523,
  title        = {Gradient-based optimisation of the conditional-value-at-risk using the multi-level Monte Carlo method},
  author       = {Sundar Ganesh and Fabio Nobile},
  year         = 2023,
  journal      = {Journal of Computational Physics},
  volume       = 495,
  pages        = 112523,
  doi          = {https://doi.org/10.1016/j.jcp.2023.112523},
  issn         = {0021-9991},
  url          = {https://www.sciencedirect.com/science/article/pii/S0021999123006186}
}

@book{smith2024,
author = {Smith, Ralph C.},
title = {Uncertainty Quantification: Theory, Implementation, and Applications, Second Edition},
publisher = {Society for Industrial and Applied Mathematics},
year = {2024},
doi = {10.1137/1.9781611977844},
address = {Philadelphia, PA},
}

@article{karve2020digital,
  author    = {Karve, Pranav M and Guo, Yulin and Kapusuzoglu, Berkcan and Mahadevan, Sankaran and Haile, Mulugeta A},
  journal   = {Engineering Fracture Mechanics},
  title     = {Digital twin approach for damage-tolerant mission planning under uncertainty},
  year      = {2020},
  pages     = {106766},
  volume    = {225},
  doi       = {10.1016/j.engfracmech.2019.106766},
  publisher = {Elsevier},
}

@article{laubenbacher2024digital,
  author    = {Laubenbacher, Reinhard and Mehrad, Borna and Shmulevich, Ilya and Trayanova, Natalia},
  journal   = {Nature Computational Science},
  title     = {Digital twins in medicine},
  year      = {2024},
  number    = {3},
  pages     = {184--191},
  volume    = {4},
  doi       = {10.1038/s43588-024-00607-6},
  publisher = {Nature Publishing Group US New York},
}

@article{sel2025survey,
  author    = {Sel, Kaan and Hawkins-Daarud, Andrea and Chaudhuri, Anirban and Osman, Deen and Bahai, Ahmad and Paydarfar, David and Willcox, Karen and Chung, Caroline and Jafari, Roozbeh},
  journal   = {npj Digital Medicine},
  title     = {Survey and perspective on verification, validation, and uncertainty quantification of digital twins for precision medicine},
  year      = {2025},
  number    = {1},
  pages     = {40},
  volume    = {8},
  doi       = {10.1038/s41746-025-01447-y},
  publisher = {Nature Publishing Group UK London},
}

@article{chaudhuripredictive2023,
  author  = {Chaudhuri, Anirban and Pash, Graham and Hormuth, David A. and Lorenzo, Guillermo and Kapteyn, Michael and Wu, Chengyue and Lima, Ernesto A. B. F. and Yankeelov, Thomas E. and Willcox, Karen},
  journal = {Frontiers in Artificial Intelligence},
  title   = {Predictive digital twin for optimizing patient-specific radiotherapy regimens under uncertainty in high-grade gliomas},
  year    = {2023},
  issn    = {2624-8212},
  month   = oct,
  pages   = {1222612},
  volume  = {6},
  doi     = {10.3389/frai.2023.1222612},
}

\end{document}